\RequirePackage[switch, columnwise, running, mathlines, displaymath
]{lineno}
\RequirePackage{docswitch}
\def\flag{mnras}
\setjournal{\flag}

\documentclass[\docopts]{\docclass}

\newcommand*\patchAmsMathEnvironmentForLineno[1]{%
  \expandafter\let\csname old#1\expandafter\endcsname\csname #1\endcsname
  \expandafter\let\csname oldend#1\expandafter\endcsname\csname end#1\endcsname
  \renewenvironment{#1}%
     {\linenomath\csname old#1\endcsname}%
     {\csname oldend#1\endcsname\endlinenomath}}%
\newcommand*\patchBothAmsMathEnvironmentsForLineno[1]{%
  \patchAmsMathEnvironmentForLineno{#1}%
  \patchAmsMathEnvironmentForLineno{#1*}}%
\AtBeginDocument{%
\patchBothAmsMathEnvironmentsForLineno{equation}%
\patchBothAmsMathEnvironmentsForLineno{align}%
\patchBothAmsMathEnvironmentsForLineno{flalign}%
\patchBothAmsMathEnvironmentsForLineno{alignat}%
\patchBothAmsMathEnvironmentsForLineno{gather}%
\patchBothAmsMathEnvironmentsForLineno{multline}%
}

\usepackage{lsstdesc_macros}
\usepackage[colorinlistoftodos]{todonotes}
\usepackage{multirow}
\usepackage{xspace}
\usepackage{enumitem}
\usepackage{graphicx}
\usepackage{wasysym}
\usepackage{amsmath}
\usepackage{amssymb}
\usepackage{amsfonts}
\usepackage{footmisc}
\usepackage{ulem}
\graphicspath{{./}{./figures/}}
\newcommand{\proj}[1]{\textsc{#1}}
\newcommand{\lsst}{\proj{LSST}}
\newcommand{\lsstdesc}{\lsst-\proj{DESC}}
\newcommand{\sdss}{\proj{SDSS}}
\newcommand{\buzz}{\proj{Buzzard}}

\newcommand{\pz}{photo-$z$}
\newcommand{\pzpdf}{\pz\ PDF}
\newcommand{\Pz}{Photo-$z$}
\newcommand{\Pzpdf}{\Pz\ PDF}
\newcommand{\chisq}{$\chi^{2}$}

\newcommand{\pzcode}[1]{\texttt{#1}}
\newcommand{\qp}{\pzcode{qp}}
\newcommand{\annz}{\pzcode{ANNz2}}
\newcommand{\bpz}{\pzcode{BPZ}}
\newcommand{\cmnn}{\pzcode{CMNN}}
\newcommand{\delight}{\pzcode{Delight}}
\newcommand{\eazy}{\pzcode{EAZY}}
\newcommand{\flexzboost}{\pzcode{FlexZBoost}}
\newcommand{\gpz}{\pzcode{GPz}}
\newcommand{\lephare}{\pzcode{LePhare}}
\newcommand{\metaphor}{\pzcode{METAPhoR}}
\newcommand{\skynet}{\pzcode{SkyNet}}
\newcommand{\tpz}{\pzcode{TPZ}}
\newcommand{\trainz}{\pzcode{trainZ}}

\definecolor{scc}{rgb}{0.0, 0.26, 0.15}
\definecolor{newpink}{rgb}{0.858, 0.188, 0.478}
\newcommand{\boldblue}[1]{{ \textcolor{black}{#1}}}
\newcommand{\xsout}[1]{}

\def\X{{\mathbf{X}}}
\def\x{{\mathbf{x}}}

\begin{document}
%\linenumbers

\title[Photo-z estimation approaches for LSST]{Evaluation of probabilistic photometric redshift estimation approaches for \boldblue{ \textsc{The Rubin Observatory Legacy Survey of Space and Time} (LSST)}}

\maketitlepre

\begin{abstract}

Many scientific investigations of photometric galaxy surveys require redshift estimates, whose uncertainty properties are best encapsulated by photometric redshift (photo-$z$) posterior probability density functions (PDFs).
A plethora of photo-$z$ PDF estimation methodologies abound, producing discrepant results with no consensus on a preferred approach.
We present the results of a comprehensive experiment comparing twelve photo-$z$ algorithms applied to mock data produced for\xsout{Large Synoptic Survey Telescope} \boldblue{The Rubin Observatory Legacy Survey of Space and Time} (\textsc{LSST}) Dark Energy Science Collaboration (\textsc{DESC}).
By supplying perfect prior information, in the form of the complete template library and a representative training set as inputs to each code, we demonstrate the impact of the assumptions underlying each technique on the output photo-$z$ PDFs.
In the absence of a notion of true, unbiased photo-$z$ PDFs, we evaluate and interpret multiple metrics of the ensemble properties of the derived photo-$z$ PDFs as well as traditional reductions to photo-$z$ point estimates.
We report systematic biases and overall over/under-breadth of the photo-$z$ PDFs of many popular codes, which may indicate avenues for improvement in the algorithms or implementations.
Furthermore, we raise attention to the limitations of established metrics for assessing photo-$z$ PDF accuracy; though we identify the conditional density estimate (CDE) loss as a promising metric of photo-$z$ PDF performance in the case where true redshifts are available but true photo-$z$ PDFs are not, we emphasize the need for science-specific performance metrics.

\end{abstract}

% Keywords are ignored in the LSST DESC Note style:
\dockeys{galaxies: distances and redshifts -- galaxies: statistics -- methods: statistical}

\maketitlepost

% ----------------------------------------------------------------------
% {% raw %}

\section{Introduction}
\label{sec:intro}

The current and next generations of large-scale galaxy surveys, including the Dark Energy Survey \citep[\proj{DES},][]{Abbott:05}, the Kilo-Degree Survey \citep[\proj{KiDS},][]{de_Jong:13}, Hyper Suprime-Cam Survey \citep[\proj{HSC},][]{Aihara:2018a,Aihara:2018b}, \xsout{Large Synoptic Survey Telescope}\boldblue{The Rubin Observatory Legacy Survey of Space and Time} \citep[\lsst,][]{Abell:09}, Euclid \citep{Laureijs:11}, and Wide-Field Infrared Survey Telescope \citep[\proj{WFIRST},][]{Green:12}, represent a paradigm shift to reliance on photometric, rather than solely spectroscopic, galaxy catalogues of substantially larger size at a cost of lacking complete spectroscopically confirmed redshifts ($z$).
Effective astrophysical inference using the catalogues resulting from these ongoing and upcoming missions, however, necessitates accurate and precise photometric redshift (\pz) estimation methodologies.

As an example, in order for \pz\ systematics to not dominate the statistical noise floor of \lsst's main cosmological sample of several $10^{9}$ galaxies, the \lsst\ Science Requirements Document (SRD)\footnote{available at \url{https://docushare.lsstcorp.org/docushare/dsweb/Get/LPM-17}} specifies that individual galaxy \pz s must have root-mean-square error $\sigma_z < 0.02 (1+z)$, $3 \sigma$ catastrophic outlier rate below $10$ per cent, and bias below $0.003$.
Specific science cases may have their own requirements on \pz\ performance that exceed those of the survey as a whole.
In that vein, the \lsst\ Dark Energy Science Collaboration (\lsstdesc) developed a separate SRD \citep{Mandelbaum:2018} that conservatively forecasts the constraining power of five cosmological probes, leading to even more stringent requirements on \pz\ performance, including those defined in terms of tomographically binned subsample populations\footnote{While tomographic samples will play a prominent role in some science cases, estimation of tomographic distributions is a distinct problem with distinct solutions. In this paper we focus solely on individual galaxy redshift estimates.} rather than individual galaxies.

Though the standard has long been for each galaxy in a photometric catalogue to have a \pz\ point estimate and Gaussian error bar, even early applications of \pz s in precision cosmology indicate the inadequacy of point estimates \citep{mandelbaum:2008} to encapsulate the degeneracies resulting from the nontrivial mapping between broad band fluxes and redshift.
Far from a hypothetical situation, such degeneracies are real consequences of the same deep imaging that enables larger galaxy catalogue sizes.
The lower luminosity and higher redshift populations captured by deeper imaging introduce major physical systematics to \pz s, among them the Lyman break/Balmer break degeneracy, that did not affect shallower large area surveys like the Sloan Digital Sky Survey \citep[\textsc{SDSS},][]{York:00} and Two Micron All Sky Survey \citep[\textsc{2MASS},][]{Skrutskie:06}.

To fully characterize such physical degeneracies, subsequent photometric galaxy catalogue data releases, \citep[e.~g.~][]{Sheldon:2012, Erben:2013, de_Jong:17}, provide a more informative \pz\ data product, the \pz\ probability density function (PDF), that describes the redshift probability, commonly denoted as $p(z)$, as a function of a galaxy's redshift, conditioned on the observed photometry.
Early template-based methods such as \citet{Fernandezsoto:99} approximated the likelihood of photometry conditioned on redshift with the relative \chisq\ values of template spectra.
Not long after, Bayesian adaptations of template-based approaches such as \citet{Benitez:00} combined the estimated likelihoods with a prior to yield a posterior PDF of redshift conditioned on photometry.
While the first data-driven \pz\ algorithms yielded a point estimate, \citet{Firth:03} estimated a \pzpdf\ using a neural net with realizations scattered within the photometric errors.  \boldblue{Several years later \citet{Wolf:09} used a data driven $\chi^{2}$ approach to estimate redshift PDFs, and specifically noted an intrinsic floor in redshift uncertainties even under ideal conditions can arise due to the intrinsic spread of redshifts within the training set even in well localized regions of parameter space.}

There are numerous techniques for deriving \pzpdf s, yet no one method has been established as clearly superior.
Consistent experimental conditions enable the quantification if not isolation of their differences, which can be interpreted as a sort of \textit{implicit prior} imparted by the method itself.
Comprehensive comparisons of \pz\ methods have been made before; the \Pz\ Accuracy And Testing \citep[\textsc{PHAT},][]{Hildebrandt:10} effort focused on \pz\ point estimates derived from many photometric bands.
\citet{Rau:2015} introduced a new method for improving \pzpdf s using an ordinal classification algorithm.
\textsc{DES} compared several codes for \pz\ point estimates and a subset with \pzpdf\ information \citep{Sanchez:14} and examined summary statistics of \pzpdf s for tomographically binned galaxy subsamples \citep{Bonnett:16}.

\boldblue{In currently available spectroscopic training sets systematic incompleteness, particularly at faint magnitudes, and incorrectly assigned redshifts are a harrowing problem.  Certain subpopulations can potentially fail to yield secure redshifts, given limited wavelength coverage.  Alternately, redshifts may be  incorrectly assigned for certain populations due to misidentification of spectral features.  These missing or misidentified populations would yield training sets that are not representative of the true underlying galaxy population, which can bias redshift estimates well beyond the levels required for upcoming analyses.  Characterizing and mitigating these problems will be a major undertaking which we will cover in future papers, as discussed briefly in \S \ref{sec:futureexperiments}.  Before taking on this very complex issue, however, we must first investigate a more fundamental one: how do current methods perform in the presence of representative data?}

This paper is distinguished from other comparisons of \pz\ methods by its focus on the evaluation criteria for \pzpdf s and interpretation thereof.
In the absence of simulated data drawn from known redshift distributions, the very concept of a ``true PDF'' for an individual galaxy is unavailable, and we must instead rely on measures of ensemble behaviour to characterize PDF quality (see \S\,\ref{sec:metrics} for further discussion).
We aim to perform a comprehensive sensitivity analysis of the dependence of \pzpdf s on the code used to produce them in order to ultimately select those that will become part of the \lsstdesc\ pipelines, described in the Science Roadmap (SRM)\footnote{Available at: \url{https://lsstdesc.org/assets/pdf/docs/DESC_SRM_latest.pdf}}.
In this initial study, we focus on evaluating the performance of \pzpdf\ codes using PDF-specific performance metrics in a formally controlled experiment with complete and representative prior information (template libraries and training sets) to set a baseline for subsequent investigations.
This approach probes how each code considered exploits the information content of the data versus prior information from template libraries and training sets.

The outline of the paper is as follows: in \S\,\ref{sec:sims} we present the simulated data set; in \S\,\ref{sec:pzcodes} we describe the current generation codes employed in the paper; in \S\,\ref{sec:metrics} we discuss the interpretation of photo-$z$ PDFs in terms of metrics of accuracy; in \S\,\ref{sec:results} we show our results and compare the performance of the codes; in \S\,\ref{sec:discussion} we offer our conclusions and discuss future extensions of this work.

\section{Data}
\label{sec:sims}

In order to test the current generation of \pzpdf\ codes, we employ an existing simulated galaxy catalogue, described in detail in Section~\ref{sec:buzzard}.
The experimental conditions shared among all codes are motivated by the \lsst\ SRD requirements and implemented for machine learning and template-based \pzpdf\ codes according to the procedures of Sections~\ref{sec:buzztraining} and \ref{sec:buzztemplates} respectively.

\subsection{The \textsc{Buzzard-v1.0} simulation}
\label{sec:buzzard}

Our mock catalogue is derived from the \textsc{Buzzard}-highres-v1.0 catalogue (DeRose et al., in prep).
\textsc{Buzzard} is built on the \texttt{Chinchilla-400} \citep{Mao:15} dark matter-only N-body simulation consisting of $2048^{3}$ particles in a $400$ Mpc h$^{-1}$ box.
The lightcone was constructed from smoothing and interpolation between a set of time snapshots.
Dark matter halos were identified using the \texttt{Rockstar} software package \citep{Behroozi:13} and then populated with galaxies with stellar masses and absolute $r$-band magnitudes in the \sdss\ system determined using a sub-halo abundance matching model constrained to match both projected two-point galaxy clustering statistics and an observed conditional stellar mass function \citep{Reddick:13}.

To assign a spectrum to each galaxy, the Adding Density Dependent Spectral Energy Distributions (SEDs) procedure \citep[\texttt{ADDSEDS,}][Wechsler et al., in prep,]{DeRose:19} was used.
\texttt{ADDSEDS} uses a sample of $\sim 5\times 10^{5}$ galaxies from the magnitude-limited \sdss\ Data Release 6 Value Added Galaxy Catalogue \citep[NYU-VAGC,][]{Blanton:05} to train an empirical relation between absolute $r$-band magnitude, local galaxy density, and SED.
Each \sdss\ spectrum is parameterized by five weights corresponding to a weighted sum of five basis SED components using the \texttt{k-correct} software package\footnote{\url{http://kcorrect.org}} \citep{Blanton:07}.

Correlations between SED and galaxy environment were included so as to preserve the colour-density relation of galaxy environments.
The distance to the spatially projected fifth-nearest neighbour was used as a proxy for local density in the \sdss\ training sample.
For each simulated galaxy, a galaxy with similar absolute $r$-band magnitude and local galaxy density was chosen from the training set, and that training galaxy's SED was assigned to the simulated galaxy.

\subsubsection{Caveats}
\label{sec:buzzlimitations}

By necessity, \textsc{Buzzard} does not contain all of the complicating factors present in real data, and here we discuss the most pertinent ways that these limitations affect our experiment.
\textsc{Buzzard} includes only galaxies, not stars nor AGN.
The catalogue-based construction excludes image-level effects, such as deblending errors, photometric measurement issues, contamination from sky background (Zodiacal light, scattered light, etc.), lensing magnification, and Galactic reddening.

The \textsc{Buzzard} SEDs are drawn from a set of $\sim 5 \times 10^{5}$ SEDs, which themselves are derived from a five-component linear combination fit to $\sim 5 \times 10^{5}$ \sdss\ galaxies; thus the sample contains only galaxies that resemble linear combinations of those for which \sdss\ obtained spectra, and there are necessarily duplicates.
The linear combination SEDs also restrict the properties of the galaxy population to linear combinations of the properties corresponding to five basis templates, precluding the modeling of non-linear features such as the full range of emission line fluxes relative to the continuum.
The only form of intrinsic dust reddening comes from what is already present in the five basis SEDs via the training set used to create the basis templates, and linear combinations thereof do not span the full range of realistic dust extinction observed in galaxy populations.

While these idealized conditions limit the realism of our mock data, they are irrelevant to the controlled experimental conditions of this study, if anything assuring that differentiation in the performance of the \pzpdf\ codes is due to the inferential techniques rather than nuances in the data.

\subsection{\lsst-like mock observations}
\label{sec:observations}

Given the SED, absolute $r$-band magnitude, and true redshift of each simulated galaxy, we computed apparent magnitudes in the six \lsst\ filter passbands, $ugrizy$.
We assigned magnitude errors in the six bands using the simple model of \citet{Ivezic:08}, assuming achievement of the full 10-year depth, with a modification of fiducial \lsst\ total numbers of 30-second visits for photometric error generation: we assume 60 visits in $u$-band, 80 visits in $g$-band, 180 visits in $r$-band, 180 visits in $i$-band, 160 visits in $z$-band, and 160 visits in $y$-band.

As a consequence of adding Gaussian-distributed photometric errors, 2.0 per cent of our galaxies exhibit a negative flux in one or more bands, the vast majority of which are in the $u$-band.
We deem such negative fluxes \textit{non-detections} and assign a placholder magnitude of 99.0 in the catalogue to indicate to the \pzpdf\ codes that such galaxies would be ``looked at but not seen'' in multi-band forced photometry.

The full dataset thus covers $400$ square degrees and contains $238$ million galaxies of redshift $0 < z \leq 8.7$ down to $r = 29$.  \boldblue{However, a problem with the k-correction calculation led to incorrect colors at $z>2$, which necessitated limiting the redshift range to $0 < z \leq 2.0$.}
\xsout{Systematic inconsistencies with galaxy colors at $z > 2$ were observed, so the catalogue was limited to $0 < z \leq 2.0$.}
To obtain a catalogue matching the \lsst\ Gold Sample, we imposed a cut of $i < 25.3$, which gives a signal-to-noise ratio $\gtrsim 30$ for most galaxies.
In order for statistical errors to be subdominant to the systematic errors we aim to probe, we further reduced the sample size to $<10^{7}$ galaxies by isolating $\sim 16.8$ square degrees selected from five separate spatial regions of the simulation.
We refer to this final set of galaxies as DC1, for the first \lsstdesc\ Data Challenge.

\subsection{Shared prior information}
\label{sec:controlled}

For the purpose of performing a controlled experiment that compares \pzpdf\ codes on equal footing as a baseline for a future sensitivity analysis, we take care to provide each with optimal prior information.
Redshift estimation approaches built upon physical modeling and machine learning alike have a notion of prior information considered beyond the photometry of the data for which redshift is to be constrained: that information is derived from a template library for a model-based code and a training set for a data-driven code.
In this initial study, we seek to set a baseline for a later comparison of the performance of \pzpdf\ codes under incomplete and non-representative prior information that will propagate differently in the space of data-driven and model-based algorithms.
However, for the baseline case of perfect prior information, physical modeling and machine learning codes can indeed be put on truly equal footing.
We outline the equivalent ways of providing all codes perfect prior information below.

\subsubsection{Training and test set division}
\label{sec:buzztraining}

Following the findings of \citet{Bernstein:10}, \citet{Newman:2015}, and \citet{Masters:2015} that\boldblue{, in the idealized case of representative training data,} only $\sim\!10^{4}$ spectra are necessary to calibrate \pz s to Stage IV requirements, we aimed to set aside a randomly selected training set of $3-5\times 10^{4}$ galaxies, $\sim 10$ per cent of the full sample.
After all cuts described above, we designated the \textit{DC1 training set} of $44\,404$ galaxies for which observed photometry, true SEDs, and true redshifts would be provided to all codes and the blinded \textit{DC1 test set} of $399\,356$ galaxies for which photometry alone would be provided to all codes and \pzpdf s would be requested.
The exact form of \lsst\ photometric filter transmission curves were also considered public information that could be used by any code.

\subsubsection{Template library construction}
\label{sec:buzztemplates}

We aimed to provide template-fitting codes with complete yet manageable library of templates spanning the space of SEDs of the DC1 galaxies.
We constructed $K=100$ representative templates from the $\sim 5 \times 10^{5}$ SEDs of the \sdss\ DR6 NYU-VAGC by using the five-dimensional vectors of SED weight coefficients described above.
After regularizing the SED weight coefficients $\in [0, 1]$, we ran a simple K-means clustering algorithm on the five-dimensional space of regularized SED weight coefficients of the \sdss\ galaxy sample.
The resulting clusters were used to define Voronoi cells in the space of weight coefficients, with centre positions corresponding to weights for the \texttt{k-correct} SED components, yielding the 100 SEDs that comprise the \textit{DC1 template set} provided to all template-based codes.
We did not, however, exclude from consideration template-based codes that made modifications in their use of these templates due to architecture limitations (as opposed to knowledge of the experimental conditions that could artificially boost the code's apparent performance), with deviations noted in Section~\ref{sec:pzcodes}.

\section{Methods}
\label{sec:pzcodes}

Here we summarize the twelve \pzpdf\ codes compared in this study, listed in Table~\ref{tab:list_of_codes}, which include both established and emerging approaches in template fitting and machine learning.
Though not exhaustive, this sample represents codes for which there was sufficient expertise within the \lsstdesc\ Photometric Redshifts Working Group.
Some aspects of data treatment were left to the individual code runners, for example, whether/how to split the available data with known redshifts into separate training and validation sets.

\xsout{Another key difference is the treatment of non-detections in one or more bands: some codes ignore incomplete bands, while others replace the value with either an estimate for the detection limit, the mean of other values in the training set, or another default value.
There are varying conventions among machine learning-based codes for treatment of non-detections, and no one prescription dominates in the \pz\ literature.
However, we remind the reader that only 2.0 per cent of our sample has non-detections, almost exclusively in the $u$-band, and thus should not dominate the code performance differences.}

\begin{table*}  %%% DATA TABLE %%%
\caption{List of \pzpdf\ codes featured in this study} \label{tab:list_of_codes}\resizebox{\textwidth}{!}{
\begin{tabular}{lll}
\hline
\bf Published code & \bf Type & \bf Public source code \\
\hline
\bpz~\citep{Benitez:00}    & template fitting& \url{http://www.stsci.edu/~dcoe/BPZ/} \\
\eazy~\citep{Brammer:08}   & template fitting & \url{https://github.com/gbrammer/eazy-photoz} \\
\lephare~\citep{Arnouts:99}   & template fitting& \url{http://www.cfht.hawaii.edu/~arnouts/lephare.html} \\
\annz~\citep{Sadeh:16}     & machine learning& \url{https://github.com/IftachSadeh/ANNZ} \\
\cmnn~\citep{Graham:17}        & machine learning & \url{https://github.com/dirac-institute/CMNN_Photoz_Estimator} \\
\delight~\citep{Leistedt:17}   & hybrid           & \url{https://github.com/ixkael/Delight} \\
\flexzboost~\citep{Izbicki:17} & machine learning & \url{https://github.com/tpospisi/flexcode}; \url{https://github.com/rizbicki/FlexCoDE}\\
\gpz~\citep{Almosallam:15b}   & machine learning& \url{https://github.com/OxfordML/GPz} \\
\metaphor~\citep{Cavuoti:17}   & machine learning& \url{http://dame.dsf.unina.it}\\
\skynet~\citep{Graff:14}       & machine learning & \url{http://ccpforge.cse.rl.ac.uk/gf/project/skynet/} \\
\tpz~\citep{Carrasco_Kind:13} & machine learning& \url{https://github.com/mgckind/MLZ} \\
\hline
\trainz                              & --& See Section~\ref{sec:trainz} \\
\end{tabular}}
\end{table*}

We describe the algorithms and implementations of the model-based and data-driven codes in Sections~\ref{sec:templatecodes} and \ref{sec:trainingcodes} respectively, with a straw-person approach included in Section~\ref{sec:trainz}.

\subsection{Template-based Approaches}
\label{sec:templatecodes}

We test three publicly available and commonly used template-based codes that share the standard physically motivated approach of calculating model fluxes for a set of template SEDs on a grid of redshift values and evaluating a \chisq\ merit function using the observed and model fluxes:
\begin{equation} \label{eq_temp_chi}
\chi^{2}(z,T,A) = \sum_{i}^{N_{\mathrm{filt}}}\left(\frac{F^{i}_{\mathrm{obs}} - A \, F^{i}_{\mathrm{pred}}(T,z)}{\sigma^{i}_{\mathrm{obs}}}\right)^2
\end{equation}
\noindent where $A$ is a normalization factor, $F^i_{\mathrm{pred}}(T,z)$ is the flux predicted for a template $T$ at redshift $z$, $F^i_{\mathrm{obs}}$ is the observed flux in a given band $i$, $\sigma^i_{\mathrm{obs}}$ is the observed flux error, and $N_{\mathrm filt}$ is the total number of filters, in our case the six $ugrizy$ LSST filters.  \boldblue{All three template-based codes replace per-band non-detections with an estimate of the 1$\sigma$ magnitude limit as estimated from the photometric error model.}
\xsout{The likelihood is a sum of observed flux error $\sigma_{b}^{\mathrm{obs}}$-weighted squared differences between the observed flux $F^{\mathrm{obs}}_{b}$ and the normalized predicted flux $F^{\mathrm{mod}}_{b}(T, z)$ in $N_{\mathrm{filt}}$ photometric filters $b$, which are the \lsst\ $ugrizy$ filters in this case.}
Specific implementation details of each code, e.~g.~prior form and implementation, are described below.

\subsubsection{BPZ}
\label{sec:BPZ}
%(Sam Schmidt)

Bayesian Photometric Redshift \citep[\bpz,][]{Benitez:00} determines the likelihood $p(C \vert z, T)$ of a galaxy's observed colours $C$ for a set of SED templates $T$ at redshifts $z$.
The \bpz\ likelihood is related to the \chisq\ likelihood by $p(C \vert z, T) \propto \exp[- \chi^{2} / 2]$.
Given a Bayesian prior $p(z, T \vert m_{0})$ over apparent magnitude $m_0$ and type $T$, and assuming that the SED templates are spanning and exclusive, \bpz\ constructs the redshift posterior $p(z \vert C, m_0)$ by marginalizing over all SED templates with the form $p(z|C,m_0)\,\propto\, \sum_{T}p(C|z,T)\,p(z,T|m_0)$ \citep[Eq.~3 from][]{Benitez:00}, corresponding to setting the parameter \texttt{PROBS\_LITE=TRUE} in the \bpz\ parameter file.
The \bpz\ prior is the product of an SED template proportion that varies with apparent magnitude $p(T \vert m_{0})$ and a prior $p(z \vert T, m_{0})$ over the expected redshift as a function of apparent magnitude and SED template.
We anticipate \bpz\ to outperform other template-based approaches due to the prior that both comprehensively accounts for SED type and is calibrated to the training set.

Here we test \bpz-v 1.99.3 \citep{Benitez:00} with the DC1 template set of Section~\ref{sec:buzztemplates}.
To keep the number of free parameters manageable, the DC1 template set is pre-sorted by the rest-frame $u-g$ colour and split into three broad classes of SED template, equivalent to the E, Sp and Im/SB types.
The Bayesian prior term $p(T \vert m_{0})$ was derived directly from the DC1 training set, and the other term $p(z \vert T, m_{0})$ was chosen to be the best fit for the eleven free parameters from the functional form of \citet{Benitez:00}.
We use template interpolation, creating two linearly interpolated templates between each basis SED (sorted by rest-frame $u-g$ colour) by setting the parameter \texttt{INTERP=2}.
Prior to running the code, the non-detection placeholder magnitude was replaced with an estimate of the 1-$\sigma$ detection limit for the undetected band as a proxy for a value close to the estimated sky noise threshold.

\subsubsection{EAZY}
\label{sec:eazy}

Easy and Accurate Photometric Redshifts from Yale \citep[\eazy,][]{Brammer:08} extends the basic \chisq\ fit procedure that defines template-fitting approaches.
The algorithm models the observed photometry with a linear combination of template SEDs at each redshift.
The best-fit SED at each redshift is found by simultaneously fitting one, two, or all of the templates via \chisq\ minimization, which is distinct from marginalizing across all templates.
The minimized \chisq\ likelihood at each redshift is then combined with an apparent magnitude prior to obtain the redshift posterior PDF.
We note that the utilization of the best-fit SED conditioned on redshift rather than a proper marginalization does not lead to the correct posterior distribution, an implementation issue that has now been identified and will be addressed by the developers in the future.

In contrast with \bpz, \eazy's apparent magnitude prior is independent of SED, though it was derived empirically from the DC1 training set.
The \eazy\ architecture cannot accept a template set other than the same five basis templates employed by \texttt{k-correct} when constructing the DC1 catalogue, but, for consistency with the experimental scope of perfect prior information, \eazy's flexible \texttt{all-templates} mode was used to fit the photometric data with a linear combination of the five basis templates.
Though \eazy\ can account for uncertainty in the template set by adding in quadrature to the flux errors an empirically derived template error as a function of redshift, we set the template error to zero since the same templates were in fact used to produce the DC1 photometry.

\subsubsection{LePhare}
\label{sec:lephare}

Photometric Analysis for Redshift Estimate \citep[\lephare,][]{Arnouts:99,Ilbert:06} uses the $\chi^2$ of Equation~\ref{eq_temp_chi} to match observed colours with those predicted from a template set.
The template set can be semi-empirical or entirely synthetic.
The reported \pzpdf\ is an arbitrary normalization of the likelihood evaluated on the output redshift grid.

Here we use \lephare-v 2.2 with the DC1 template set of Section~\ref{sec:buzztemplates}.
Unlike both \bpz\ and \eazy, \lephare\ uses generic, SED-independent priors that are not tuned to the DC1 data set.  \boldblue{The choice to use the generic prior form could lead to a degradation in code performance, given the slight mismatch between the galaxy distributions in the studied dataset and that parameterized in the \lephare\ prior.}

\subsection{Machine Learning-based Approaches}
\label{sec:trainingcodes}

We compared nine data-driven \pz\ estimation approaches, eight of which are described in this section and one of which is discussed in Section~\ref{sec:trainz}.
Because the algorithms differ more from one another and the techniques are relative newcomers to the astronomical literature, we provide somewhat more detail about the implementations below.
\boldblue{A key difference between the codes is the treatment of non-detections in one or more bands: some codes ignore per-band non-detections, while others replace the value with either an estimate for the detection limit, the mean of other values in the training set, or another default value.
There are varying conventions among machine learning-based codes for treatment of non-detections, and no one prescription dominates in the \pz\ literature.
However, we remind the reader that only 2.0 per cent of our sample has non-detections, almost exclusively in the $u$-band, and thus should not dominate the code performance differences.
}

\subsubsection{ANNz2}
\label{sec:annz2}

\annz\ \citep{Sadeh:16} supports several machine learning algorithms, including artificial neural networks (ANN), boosted decision tree, and k-nearest neighbour (KNN) regression.
In addition to accounting for errors on the input photometry, \annz\ uses the KNN-uncertainty estimate of \citet{Oyaizu:08} to quantify uncertainty in the choice of method over multiple runs.
Using the Toolkit for Multivariate Data Analysis with ROOT\footnote{\url{http://tmva.sourceforge.net/}}, \annz\ can return the results of running a single machine learning algorithm, a ``best'' choice of the results from simultaneously running multiple algorithms (based on evaluation the cumulative distribution functions of validation set objects), or a combination of the results of multiple algorithms weighted by their method uncertainties averaged over multiple runs.

In this study, we use \annz-v.2.0.4.  \boldblue{Several combinations of the ANN, BDT, and kNN machine learning algorithms were tested during the training/validation stage.  For this particular dataset, the optimal and most stable results came from a setup consisting solely of ANNs.  We present results for this case and do not include results from a mixture of multiple algorithms. }
\xsout{to output only the result of the ANN algorithm.}
\Pzpdf s were produced by running an ensemble of 5 ANNs with a $6:12:12:1$ architecture corresponding to the 6 $ugrizy$ inputs, 2 hidden layers with 12 nodes each, and 1 output of redshift.
Each of the five ANNs was trained with different random seeds for the initialization of input parameters, reserving half of the training set for validation to prevent overfitting.
\boldblue{Galaxies not detected in the $u$-band} \xsout{Undetected galaxies}were excluded from the training set, and per-band non-detections in the test set were replaced with the mean magnitude \boldblue{of the test set} in that band. \xsout{within the entire test set.}

\subsubsection{Colour-Matched Nearest-Neighbours}
\label{sec:cmnn}

The colour-matched nearest-neighbours photometric redshift estimator \citep[\cmnn,][]{Graham:17} uses a training set of galaxies with known redshifts that has equivalent or better photometry than the test set in terms of quality and filter coverage.
For each galaxy in the test set, \cmnn\ identifies a colour-matched subset of training galaxies using a threshold in the Mahalanobis distance $D_M = \sum_{j}^{N_{\mathrm{colours}}} (c_{\mathrm{j,train}} - c_{\mathrm{j,test}})^{2} / \delta c_{\mathrm{j,test}}^2$ in the space of available colours $c$, with colour measurement errors $\delta c_{\mathrm{test}}$ and $N_{\mathrm{colours}} = 5$ colours $j$ defined by the $ugrizy$ filters, which defines the set of colour-matched neighbours based on a value of the percent point function (PPF).
As an example, for $N_{\mathrm{filt}}=5$ with PPF$=0.95$, $95$ per cent of all training galaxies consistent with the test galaxy will have $D_M < 11.07$.
\xsout{Undetected bands}\boldblue{Bands in which the galaxy is not detected} are dropped, thereby reducing the effective $N_{\mathrm{filt}}$ for that galaxy.
The \pzpdf\ of a given test set galaxy is the normalized distribution of redshifts of its colour-matched subset of training set galaxies.

Here, we make two modifications to the implementation of \citet{Graham:17} to comply with the controlled experimental conditions.
First, we do not impose non-detections on galaxies fainter than the expected \lsst\ 10-year limiting magnitude nor galaxies bright enough to saturate with \lsst's CCDs, instead using all of the photometry for the DC1 test and training sets.
Second, we apply the initial colour cut to the training set before calculating the Mahalanobis distance in order to accelerate processing and use a magnitude pseudo-prior as in \citet{Graham:17}, but for both we use cut-off values corresponding to the DC1 training set galaxies' colours and magnitudes.

We make an additional adaptation to enable the \cmnn\ algorithm to yield accurate \pzpdf s for all galaxies, as the original \citet{Graham:17} algorithm is optimized for \pz\ point estimates and is susceptible to less accurate \pzpdf s for bright galaxies or those with few matches in colour-space.
We use PPF$=0.95$ rather than PPF$=0.68$ to generate the subset of colour-matched training galaxies, whose redshifts are weighted by their inverse Mahalanobis distances when composing the \pzpdf\ rather than weighting all colour-matched training galaxies equally.
Additionally, when the number of colour-matched training set galaxies is less than 20, the nearest 20 neighbours in colour-space are used instead, and the output \pzpdf\ is convolved with a Gaussian kernel of variance $\sigma_{\mathrm{ train}}^{2}(\mathrm{PPF}_{20}/0.95)^2 -1$ to account for the corresponding growth of the effective PPF to include 20 neighbours.

\subsubsection{Delight}
\label{sec:delight}
%(John Soo)

\delight\ \citep{Leistedt:17} is a hybrid technique that infers \pz s with a data-driven model of latent SEDs and a physical model of photometric fluxes as a function of redshift.
Generally, machine learning methods rely on representative training data with shared photometric filters, while template-based methods rely on a complete library of templates based on physical models constructed.
\delight\ aims to take the best aspects of both approaches by constructing a large collection of latent SED templates (or physical flux-redshift models) from training data, with a template SED library as a guide to the learning of the model, thereby circumventing the machine learning prerequisite of representative training data in the same photometric bands and the template fitting requirement of detailed galaxy SED models.
It models noisy observed flux $\mathbf{\hat{F}} = \mathbf{F} + F_{b}$ as a sum of a noiseless flux plus a Gaussian processes $F_b \sim \mathcal{GP}\left(\mu^F, k^F \right)$ with zero mean function $\mu^{F}$ and a physically motivated kernel $k^{F}$ that induces realistic correlations in flux-redshift space.

From a template-fitting perspective, each test set galaxy has a posterior $p(z \vert \mathbf{\hat{F}}) \approx \sum_i p(\mathbf{\hat{F}} \vert z, T_i) p(z \vert T_i) p(T_i)$ of redshift $z$ conditioned on noisy flux $\mathbf{\hat{F}}$, where $p(z \vert T_i) p(T_i)$ captures prior information about the redshift distributions and abundances of the galaxy templates $T_i$.
As in traditional template fitting, each likelihood $p(\mathbf{\hat{F}} \vert \mathbf{F})$ relates the noisy flux $\mathbf{\hat{F}}$ with the noiseless flux $\mathbf{F}$ predicted by the model of a linear combination of templates, carefully constructed to account for model uncertainties and different normalization of the same SED, plus the Gaussian process term.

The machine learning approach appears in the inclusion of a pairwise comparison term $p(\mathbf{F} \vert z, z_j, \mathbf{\hat{F}}_j)$ for the prediction of model flux $\mathbf{F}$ at a model redshift $z$ with respect to training set galaxy $j$ with redshift $z_j$ and observed flux $\mathbf{\hat{F}}_j$.
Thus the \pz\ posterior $p(\mathbf{\hat{F}} \vert z, T_i) = \int p(\mathbf{\hat{F}} \vert \mathbf{F}) p(\mathbf{F} \vert z, z_j, \mathbf{\hat{F}}_j) d\mathbf{F}$ may be interpreted as the probability that the training and the target galaxies have the same SED at different redshifts.
The flux prediction $p(\mathbf{F} \vert z, z_j, \mathbf{\hat{F}}_j)$ of the training galaxy at redshift $z$ is modeled via the Gaussian process described above; more detail is provided in \citet{Leistedt:17}.

In this study, the default settings of \delight\ were used, with the exception that the PDF bins were set to be linearly spaced rather than logarithmically.
The Gaussian process was trained using the full DC1 training set.
We used the full DC1 template set with a flat prior in magnitude and SED type.
Photometric uncertainties from the inputs are propagated into the code, while non-detections for each band are set to the mean of the respective bands.

\subsubsection{FlexZBoost}
\label{sec:flexzboost}

\flexzboost\ \citep{Izbicki:17,Dalmasso:2019} is built on \texttt{FlexCode}, a general-purpose methodology for converting any conditional mean point estimator of $z$ to a conditional density estimator $p(z \vert \x) \equiv f(z \vert \x)$, where $\x$ here represents our photometric covariates and errors.
\flexzboost\ expands the unknown function $f(z \vert \x) = \sum_{i}\beta_{i}(\x)\phi_{i}(z)$ using an orthonormal basis $\{\phi_{i}(z)\}_{i}$.
By the orthogonality property, the expansion coefficients $\beta_{i}(\x) = \mathbb{E}\left[\phi_i(z)|\x\right] \equiv \int f(z \vert \x) \phi_{i}(z) dz$ are thus conditional means.
The expectation value $\mathbb{E}\left[\phi_i(z) \vert \x\right]$ of the expansion coefficients conditioned on the data is equivalent to the regression of the space of possible redshifts on the space of possible photometry.
Thus the expansion coefficients $\beta_{i}(\x)$ can be estimated from the data via regression to yield the conditional density estimate $\widehat{f}(z \vert \x)$.

In this paper, we used \texttt{xgboost} \citep{Chen:16} for the regression; it should, however, be noted that \texttt{FlexCode-RF}\footnote{\url{https://github.com/tpospisi/flexcode};\\ \url{https://github.com/rizbicki/FlexCoDE} \label{flexzboost_github}}, based on Random Forests, generally performs better on smaller datasets.
As our basis $\phi_{i}(z)$, we choose a standard Fourier basis.
The two tuning parameters in our \pzpdf\ estimate are the number $I$ of terms in the series expansion and an exponent $\alpha$ that we use to sharpen the computed density estimates $\widetilde{f}(z \vert \x) \propto \widehat{f}(z \vert \x)^{\alpha}$.
Both $I$ and $\alpha$ were chosen in an automated way by minimizing the weighted $L_2$-loss function \citep[Eq. 5 in][]{Izbicki:17} on a validation set comprised of a randomly selected 15 per cent of the DC1 training set.
While \texttt{FlexCode}'s lossless native encoding stores each \pzpdf\ using the basis coefficients $\beta_{i}(\x)$, we discretized the final estimates into 200 linearly spaced redshift bins $0 < z < 2$ to match the consistent output format of the experimental conditions.

\subsubsection{GPz}
\label{sec:gpz}

\gpz\ \citep{Almosallam:16a,Almosallam:15b} is a sparse Gaussian process-based code, a scalable approximation of full Gaussian Processes \citep{Rasmussen:06}, that produces input-dependent variance estimates corresponding to heteroscedastic noise.
The model assumes a Gaussian posterior probability $p(z \vert \x) = \mathcal{N}\left(z \vert \mu(\x), \sigma(\x)^{2}\right)$ of the output redshift $z$ given the input photometry $\x$.
The mean $\mu(\x)$ and the variance $\sigma(\x)^{2}$ are modeled as functions $f(\x) = \sum_{i=1}^{m}w_{i}\phi_{i}(\x)$ that are linear combinations of $m$ basis functions $\left\{\phi_{i}(\x)\right\}_{i=1}^{m}$ with associated weights $\left\{w_{i}\right\}_{i=1}^{m}$.
The details on how to learn the parameters of the model and the hyper-parameters of the basis functions are described in \citet{Almosallam:15b}.
\gpz's variance estimate is composed of a model uncertainty term corresponding to sparsity of the training set photometry and a noise uncertainty term encompassing noisy photometric observations, enabling quantification of any need for more representative or more precise training samples.
\gpz\ may also weight training set samples by importance according to $|z_{\mathrm{spec}} - z_{\mathrm{phot}}| / (1+z_{\mathrm{spec}})$ to minimize the normalized \pz\ point estimate error.  However, this function may be adapted to \pzpdf s, adding weight to test set galaxies that are not well-represented in the training set.

To smooth the long tail in the distribution of magnitude errors, we use the logarithm of the magnitude errors, improving numerical stability and eliminating the need for constraints on the optimization process.
Unobserved magnitudes $x_{\mathrm{u}} = \mu_{\mathrm{u}} + \Sigma_{\mathrm{uo}}\Sigma_{\mathrm{oo}}^{-1}(x_{\mathrm{o}} - \mu_{\mathrm{o}})$ were imputed from observed magnitudes $x_{\mathrm{o}}$ and the training set mean $\mu$ and covariance $\Sigma$ using a linear model.
This is the optimal expected value of the unobserved variables given the observed ones under the assumption that the distribution is jointly Gaussian; note that this reduces to a simple average if the covariates are independent with $\Sigma_{\mathrm{uo}} = 0$.
We reserved for validation 20 per cent of the training set and used the Variable Covariance option in \gpz\ with 200 basis functions (see \citet{Almosallam:15b} for details), and did not apply cost-sensitive learning options.

\subsubsection{METAPhoR}
\label{sec:metaphor}

Machine-learning Estimation Tool for Accurate Photometric Redshifts \citep[\metaphor,][]{Cavuoti:17} is based on the Multi Layer Perceptron with Quasi Newton Algorithm (MLPQNA) with the least square error model and Tikhonov $L_{2}$-norm regularization \citep{Hofmann:18}.
\Pzpdf s are generated by running $N$ trainings on the same training set, or $M$ trainings on $M$ different random samplings of the training set.
Upon regression of the test set, the photometry $m_{ij}$ of each test set galaxy $j$ in filter $i$ is perturbed according to $m_{ij}' = m_{ij} + \alpha_{i} F_{ij} \epsilon$ in terms of the standard normal random variable $\epsilon \sim \mathcal{N}(0, 1)$, a multiplicative constant $\alpha_{i}$ permitting accommodation of multi-survey photometry, and a bimodal function $F_{ij}$ composed of a polynomial fit of the mean magnitude errors on the binned bands plus a constant term representing the threshold below which the polynomial's noise contribution is negligible \citep{Brescia:18}.

In this work, we used a hierarchical KNN to replace non-detections with values based on their neighbours.
The usual cross-validation of redshift estimates and PDFs was also omitted for this study.

\subsubsection{SkyNet}
\label{sec:skynet}

\skynet\ \citep{Graff:14} employs a neural network based on a second-order conjugate gradient optimization scheme \citep[see][for further details]{Graff:14}. 
The neural network is configured as a standard multilayer perceptron with three hidden layers and one input layer with 12 nodes corresponding to the 6 photometric magnitudes and their measurement errors.

\skynet's classifier mode uses a cross-entropy error function with a 20:40:40 node (all rectified linear units) architecture for each hidden layer and an output layer of 200 nodes corresponding to 200 bins for the PDF, with a softmax activation function to enforce the normalization condition that the probabilities sum to unity.
While previous implementations of the code \citep[see Appendix C.3 of~][]{Sanchez:14,Bonnett:15} implement a sliding bin smoothing, no such procedure was used in this study.

We pre-whitened the data by pegging the magnitudes to (45,45,40,35,42,42) and errors to (20,20,10,5,15,15) for $ugrizy$ filters, respectively.
To avoid over-fitting, $30$ per cent of the training set was reserved for validation, and training was halted as soon as the error rate began to increase on the validation set.
The weights were randomly initialized based on normal sampling.

\subsubsection{TPZ}
\label{sec:tpz}

Trees for \Pz\ \citep[\tpz,][]{Carrasco_Kind:13,Carrascokind:14} uses prediction trees and random forest techniques to estimate \pzpdf s.
\tpz\ recursively splits the training set into branch pairs based on maximizing information gain among a random subsample of features, to minimize correlation between the trees, terminating only when a newly created leaf meets a criterion, such as a leaf size minimum or a variance threshold.
The regions in each terminal leaf node correspond to a subsample of the training set with similar properties.
Bootstrap samples from the training set photometry and errors are used to build a set of prediction trees.

To run \tpz, we replaced non-detections with an approximation of the $1\sigma$ detection threshold based on the signal-to-noise-based error forecast of the 10-year \lsst\ data, i.~e.~$dm = 2.5 \log (1 + N/S)$ where $dm \sim 0.7526$ magnitudes for $N/S = 1$ (where $N$ and $S$ are the noise and signal).
We calibrated \tpz\ with the Out-of-Bag cross-validation technique \citep{Breiman:84,Carrasco_Kind:13} to evaluate its predictive validity and determine the relative importance of the different input attributes.
We grew 100 trees to a minimum leaf size of 5 using the $ugri$ magnitudes, all $u-g, g-r, r-i, i-z, z-y$ colours, and the associated errors, as the $z$ and $y$ magnitudes did not show significant correlation with the redshift in our cross-validation.
We partitioned our redshift space into 200 bins and smoothed each individual PDF with a smoothing scale of twice the bin size.

\subsection{trainZ: a pathological \pzpdf\ estimator}
\label{sec:trainz}

We also consider a pathological \pzpdf\ estimation method, dubbed \trainz, which assigns each test set galaxy a \pzpdf\ equal to the normalized redshift distribution $N(z)$ of the training set, according to
\begin{equation}
  p(z \vert \{z_{j}\}) \equiv \frac{1}{N_{ \mathrm train}}\sum_{\mathrm i=1}^{N_{\mathrm train}} \begin{cases} 1 & \text{if\ } z_{k}\leq z_{i} < z_{k+1}\\ 0 & \text{otherwise} \end{cases}.
\end{equation}
Unlike the other methods, the \trainz\ estimator is \textit{independent of the photometric data}, effectively performing a KNN procedure with $k=N_{\mathrm{train}}$, a limiting case of a \pzpdf\ estimator dominated by the shared prior information of the training set.
In this way, \trainz\ serves as an experimental control that is not a competitive \pzpdf\ method that would be used by any real survey.

Though \trainz\ is strongly vulnerable to a nonrepresenative training set, it should optimize performance metrics probing the ensemble properties of the galaxy sample, modulo Poisson error due to small sample size, as the training set and test set are drawn from the same underlying population.
We will demonstrate its performance under the metrics of Section~\ref{sec:metrics} and discuss it as an illustrative experimental control case in Section~\ref{sec:caution} to highlight the limitations of our evaluation criteria for \pzpdf s.

\section{Analysis}
\label{sec:metrics}

The goal of this study is to evaluate the degree to which \pzpdf s of each method can be trusted for a generic analysis.
The overloaded ``$p(z)$'' is a widespread abuse of notation that obfuscates this goal, so we dedicate attention to dismantling it here.

Galaxies have redshifts $z$ and photometric data $d$ drawn from a joint probability space $p(z, d)$ in nature, and each observed galaxy $i$ has a \textit{true posterior \pzpdf}\ $p(z \vert d_{i})$.
There are a number of metrics that can be used to test the accuracy of a \pz\ posterior as an estimator of a true \pz\ posterior if the true \pzpdf\ is known.
However, the true \pzpdf\ of the observed data is not accessible, as existing mock catalogues produce redshift-photometry pairs $(z, d)$ by a deterministic algorithm that does not correspond to a joint probability density from which one can take samples.
In these cases there is no ``true PDF'' for an individual object
\footnote{\boldblue{While a discrete approximation to the true $p(z,d)$ is possible by sampling the local neighbourhood of parameter space in large datasets with methods like a nearest neighbor or conditional density estimate, we do not make an explicit computation of such an approximate distribution.  Rather, we note that the CMNN and FlexZBoost algorithms are specific implementations of such algorithms which can be used as examples of such approximations in the absence of knowledge of the true $p(z,d)$.}}, and most measures of PDF fidelity will necessarily be restricted to probing the quality of the ensemble of \pzpdf s.
(See \S,\ref{sec:futureexperiments} for a discussion of how one might circumvent this limitation.)

Before describing the metrics appropriate to the DC1 data set, we outline the philosophy behind our choices.
A \pzpdf\ estimator derived by method $H$ must be understood as a posterior probability distribution
\begin{equation}
  \label{eq:pzpdf}
\hat{p}^{H}_{i}(z) \equiv p(z \vert d_{i}, I_{D}, I_{H}),
\end{equation}
conditioned not only on the photometric data $d_{i}$ for that galaxy but also on parameters encompassing prior information $I_{D}$ shared, in our experiment, among all \pzpdf\ codes and $I_{H}$ that will differ depending on the method $H$ used to produce it.
To be concrete, $I_{D}$ takes the form of a training set for the machine learning codes and a template library for the model fitting codes.

The interpretation of the information $I_{H}$ is more subtle.
This investigation is built upon the knowledge that two codes taking the same approach, among choices of model fitting or machine learning, are nonetheless expected to yield different results even if they take the same external prior information $I_{D}$.
$I_{H}$ represents the projection of the code's architecture onto the estimated posteriors over redshift, specific to each code, and even the tunable parameters or random seeds of a specific run of a code with a random component.
We refer to $I_{H}$ as the \textit{implicit prior}, in contrast with the training set or template library provided to a given code explicitly by the researcher.  In simple terms, the implicit prior is the collection of the many different assumptions, coding choices, algorithm selections, and other implementation details that are specific to each code, the ensemble of which results in differing estimates of redshift when combined with the data and prior information in common to all codes.

The presence of the implicit prior in some sense makes a direct comparison of \pzpdf s produced by different methods impossible; even if they share the same external prior information $I_{D}$, by definition they cannot be conditioned on the same assumptions $I_{H}$, otherwise they would not be distinct methods at all.
In this study, we isolate the effect of differences in prior information $I_{H}$ specific to each method by using a single training set $I_{D}^{\mathrm{ML}}$ for all machine learning-based codes and a single template library $I_{D}^{\mathrm{T}}$ for all template-based codes.
These sets of prior information are carefully constructed to be representative and complete, so we have $I_{D} \equiv I_{D}^{\mathrm{ML}} \equiv I_{D}^{\mathrm{T}}$ for every method $H$.
Under this assumption, a ratio of posteriors of codes is in effect a ratio of the implicit posteriors $p(z \vert d_{i}, I_{H'})$ since the external prior information $I_{D}$ is present in the numerator and denominator.
Thus comparisons of $\hat{p}_{i}^{H}(z)$ isolate the effect of the method used to obtain the estimator, which should enable interpretation of the differences between estimated PDFs in terms of the specifics of the method implementations.

The exact implementation of the metrics theoretically depends on the parametrization of the \pzpdf s, which may differ across codes and can affect the precision of the estimator \citep{Malz:2018}.
Even considering a single method under the same parametrization, such as the 200-bin $0 < z < 2$ piecewise constant function used here, the exact bin definitions must affect the result.
The piecewise constant format is chosen because of its established presence in the literature, and the choice of 200 bins was motivated by the approximate number of columns expected to be available for storage of \pzpdf s for the final \lsst\ Project tables.\footnote{See, e.~g.~the \lsst\ Data Products Definition Document, available at: \url{https://ls.st/dpdd}}
We will discuss the choice of \pzpdf\ parameterization further in Section~\ref{sec:discussion}.

This analysis is conducted using the \qp\footnote{\url{http://github.com/aimalz/qp/}}\ software package \citep{Malz:qp} for manipulating and calculating metrics of univariate PDFs.
We present the metrics of \pzpdf s that address our goals in the sections below.
Section~\ref{sec:qualmet} outlines aggregate metrics of a catalogue of \pzpdf s, and Section~\ref{sec:CDE_loss} presents a metric of individual \pzpdf s in the absence of true \pzpdf s.
Those seeking a connection to previous comparison studies will find metrics of redshift point estimate reductions of \pzpdf s in Appendix~\ref{sec:pointmetrics} and metrics of a science-specific summary statistic heuristically derived from \pzpdf s in Appendix~\ref{sec:moments}.

\subsection{Metrics of \pzpdf \ ensembles}
\label{sec:qualmet}

Because \lsst's \pzpdf s will be used for many scientific applications, some of which require each individual catalogue entry to be accurate, we consider several metrics that probe the population-level performance of the \pzpdf s.
As we have the true redshifts but not true \pzpdf s for comparison, we remind the reader of the Cumulative Distribution Function (CDF)
\begin{equation}
  \label{eq:cdf}
  \mathrm{CDF}[f, q] \equiv \int_{-\infty}^{q} f(z) dz,
\end{equation}
of a generic univariate PDF $f(z)$, which is used as the basis for several of our metrics.
We describe metrics based on the CDF in Section~\ref{sec:qqpit} and metrics of summary statistics thereof in Section~\ref{sec:summqqpit}.

\subsubsection{CDF-based metrics}
\label{sec:qqpit}

A quantile of a distribution is the value $q$ at which the CDF of the distribution is equal to $Q$; percentiles and quartiles are familiar examples of linearly spaced sets of 100 and 4 quantiles, respectively.
The quantile-quantile (QQ) plot serves as a graphical visualization for comparing two distributions, where the quantiles of one distribution are plotted against the quantiles of the other distribution, providing an intuitive way to qualitatively assess the consistency between an estimated distribution and a true distribution.
The closer the QQ plot is to diagonal, the closer the match between the distributions.

The probability integral transform (PIT)
\begin{align}
\label{eq:pit}
\mathrm{PIT} &\equiv \mathrm{CDF}[\hat{p}, z_{\mathrm true}]
\end{align}
is the CDF of a \pzpdf\ evaluated at its true redshift, and the distribution of PIT values probes the average accuracy of the \pzpdf s of an ensemble of galaxies.
The distribution of PIT values is effectively the derivative of the QQ plot.
A catalogue of accurate \pzpdf s should have a PIT distribution that is uniform $U(0,1)$, and deviations from flatness are interpretable: overly broad \pzpdf s induce underrepresentation of the lowest and highest PIT values, whereas overly narrow \pzpdf s induce overrepresentation of the lowest and highest PIT values.
Catastrophic outliers with a true redshift outside the support of its \pzpdf\ have $\mathrm{PIT} \approx 0$ or $\mathrm{PIT} \approx 1$.

The PIT distribution has been used to quantify the performance of \pzpdf\ methods in the past \citep[e.~g.~][]{Bordoloi:10,Polsterer:16,Tanaka:17}.
\citet{Tanaka:17} use the histogram of PIT values as a diagnostic indicator of overall code performance, while \citet{Freeman:17} independently define the PIT and demonstrate how its individual values may be used both to perform hypothesis testing (via, e.~g.~ the KS, CvM, and AD tests; see below) and to construct QQ plots.
Following Kodra \& Newman (in prep.) we define the PIT-based catastrophic outlier rate as the fraction of galaxies with $\mathrm{PIT} < 0.0001$ or $\mathrm{PIT} > 0.9999$, which should total 0.0002 for an ideal uniform distribution.

\subsubsection{Summary statistics of CDF-based metrics}
\label{sec:summqqpit}

We evaluate a number of quantitative metrics derived from the visually interpretable QQ plot and PIT histogram, built on the Kolmogorov-Smirnov (KS) statistic
\begin{equation}
  \label{eq:ks}
  \mathrm{KS} \equiv \max_{z} \left( \left| \mathrm{CDF}[\hat{f}, z] - \mathrm{CDF}[\tilde{f}, z] \right| \right),
\end{equation}
interpretable as the maximum difference between the CDFs of the empirical distribution of PIT values for the test sample $\hat{f}(z)$ and a reference distribution $\tilde{f}(z)$, in this case $U(0,1)$, for the ideal distribution of PIT values.
We also consider two variants of the KS statistic.
A cousin of the KS statistic, the Cramer-von Mises (CvM) statistic
\begin{equation}
\label{eq:cvm}
  \mathrm{CvM}^{2} \equiv \int_{-\infty}^{+\infty} \big(\mathrm{CDF}[\hat{f}, z] - \mathrm{CDF}[\tilde{f}, z]\big)^2 \mathrm{d}\mathrm{CDF}[\tilde{f}, z]
\end{equation}
is the mean-squared difference between the CDFs of an approximate and true PDF.
The Anderson-Darling (AD) statistic
\begin{equation} \label{eq:ad}
  \mathrm{AD}^2 \equiv N_{tot}\int_{-\infty}^{+\infty} \frac{\big(\mathrm{CDF}[\hat{f}, z] - \mathrm{CDF}[\tilde{f}, z]\big)^2} {\mathrm{CDF}[\tilde{f}, z] (1 -\mathrm{CDF}[\tilde{f}, z])} \mathrm{d}\mathrm{CDF}[\tilde{f}, z]
\end{equation}
is a weighted mean-squared difference featuring enhanced sensitivity to discrepancies in the tails of the distribution.
In anticipation of a substantial fraction of galaxies having PIT of 0 or 1, a consequence of catastrophic outliers, we evaluate the AD statistic with modified bounds of integration $(0.01, 0.99)$ to exclude those extremes in the name of numerical stability.

\subsection{Conditional Density Estimate (CDE) Loss: a metric of individual \pzpdf s}
\label{sec:CDE_loss}

The \buzz\ simulation process precludes testing the degree to which samples from our \pz\ posteriors reconstruct the space of $p(z, \mathrm{data})$.
To the knowledge of the authors, there is only one metric that can be used to evaluate the performance of individual \pzpdf\ estimators in the absence of true \pz\ posteriors.
The conditional density estimation (CDE) loss is an analogue to the familiar root-mean-square-error used in conventional regression, defined as
\begin{equation}
  \label{eq:cde-loss}
  L(f, \widehat{f}) \equiv \int \int (f(z \vert \x) - \widehat{f}(z \vert \x))^{2} \mathrm{d}z \mathrm{d}P(\x) ,
\end{equation}
where $f(z \vert \x)$ is the true \pzpdf\ that we do not have and $\widehat{f}(z \vert \x)$ is an estimate thereof, in terms of the photometry $\x$.
(See Section~\ref{sec:flexzboost} for a review of the notation.)
We estimate the CDE loss via
\begin{equation}
  \label{eq:estimated-cde-loss}
  \hat{L}(f, \widehat{f}) = \mathbb{E}_\X \left[\int \widehat{f}(z \mid \X)^{2} dz\right] - 2 \mathbb{E}_{\X, Z}\left[\widehat{f}(Z \mid \X)\right] + K_{f},
\end{equation}
where the first term is the expectation value of the \pz\ posterior with respect to the marginal distribution of the photometric covariates $\X$, the second term is the expectation value with respect to the joint distribution of $\X$ and the space $Z$ of all possible redshifts, and the third term $K_{f}$ is a constant depending only upon the true conditional densities $f(z \vert \x)$\xsout{We may estimate these expectations empirically on the test or validation data} \citep[Eq.~7 in][]{Izbicki:17b}. \xsout{without knowledge of the true densities.}
\boldblue{One of the most powerful features of the CDE loss is the ability to estimate the loss function up to a constant even in the absence of knowledge of the true underlying distribution (see Eq.~7 and accompanying discussion in \citet{Izbicki:17} for more details).  This feature enables a quantitative comparison of our estimation methods with our current dataset where we lack access to the true $f(z \vert \x)$.}

\section{Results}
\label{sec:results}

\begin{figure*}
%\centering
\includegraphics[width=0.49\textwidth]{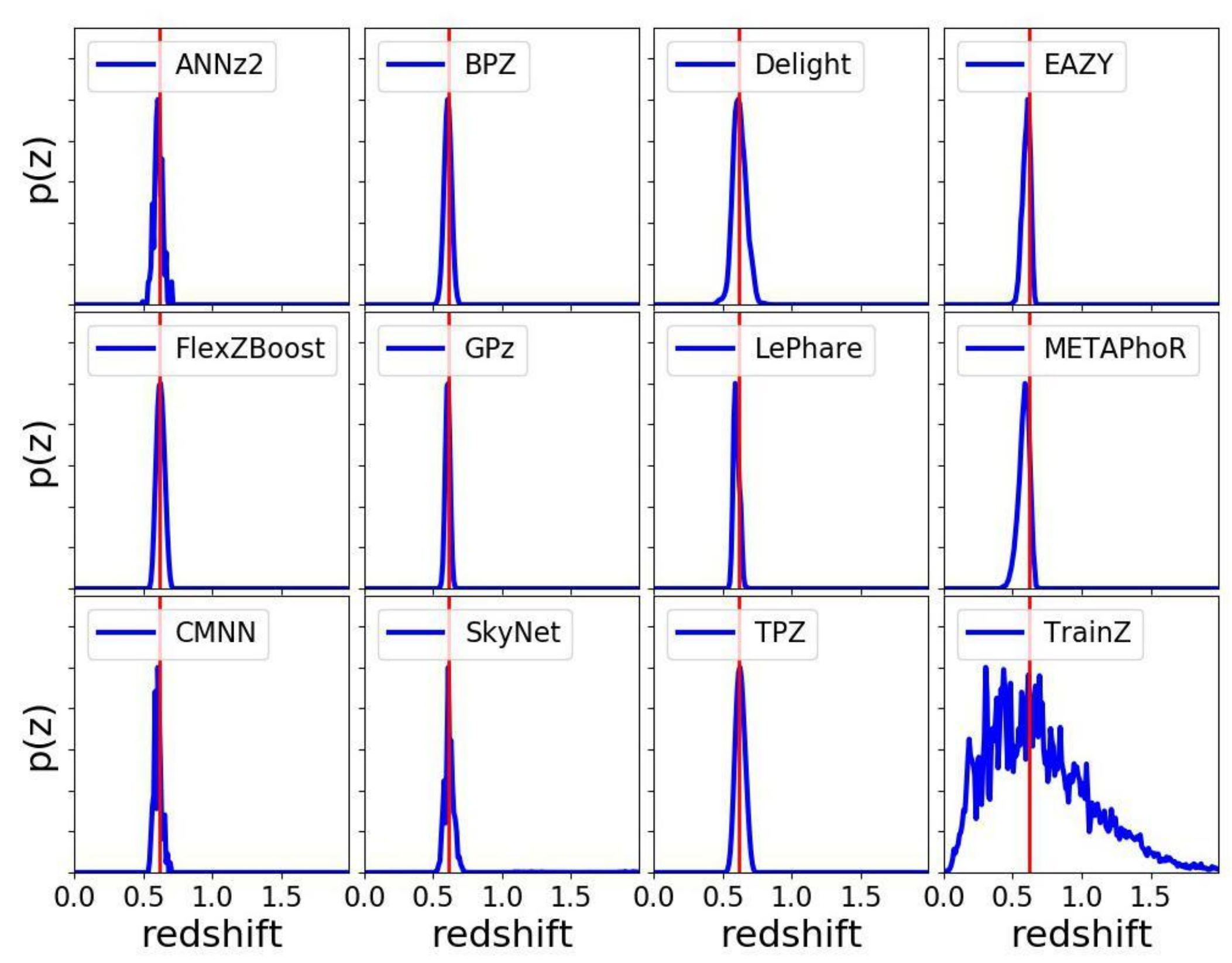}\includegraphics[width=0.49\textwidth]{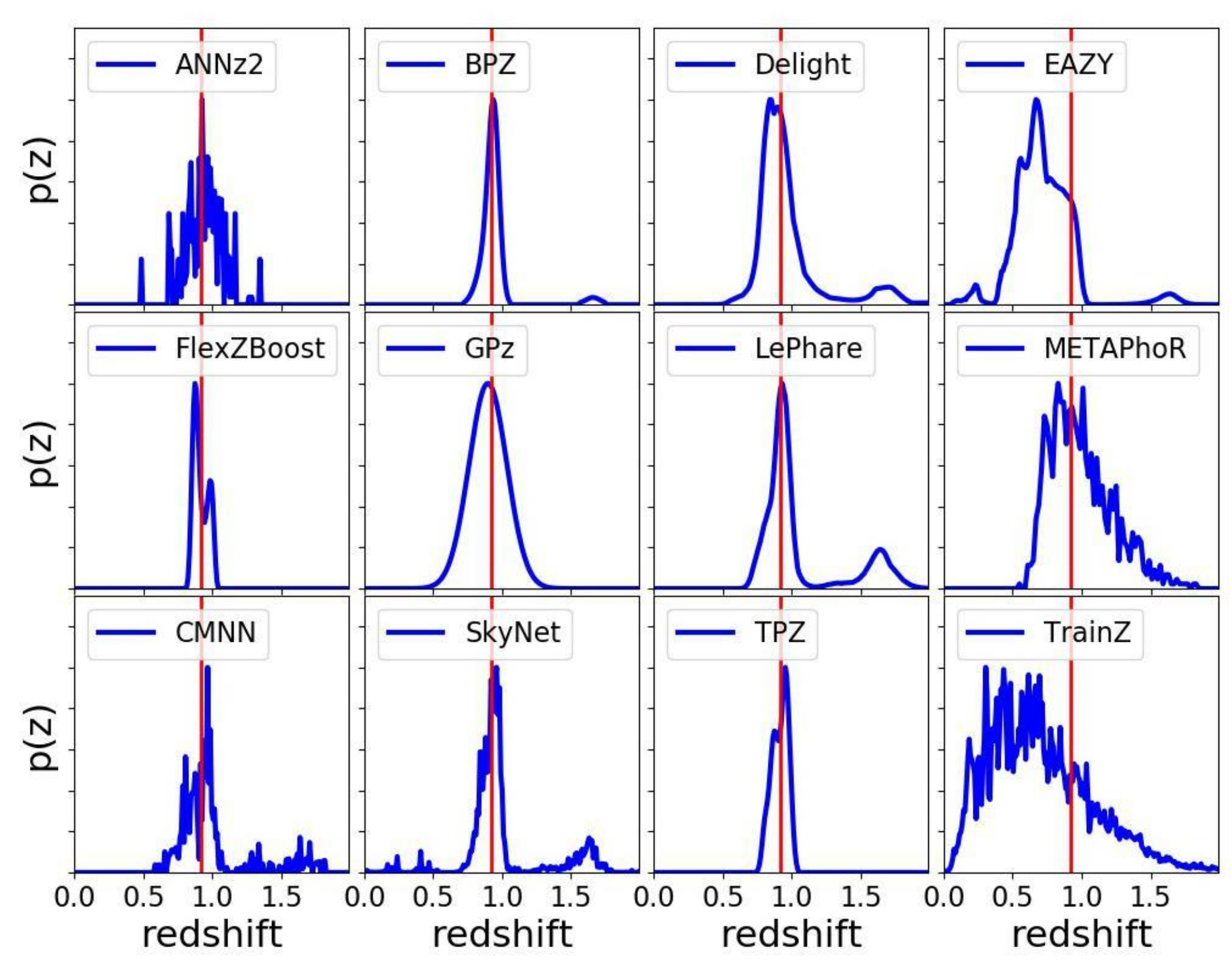}\\
\includegraphics[width=0.49\textwidth]{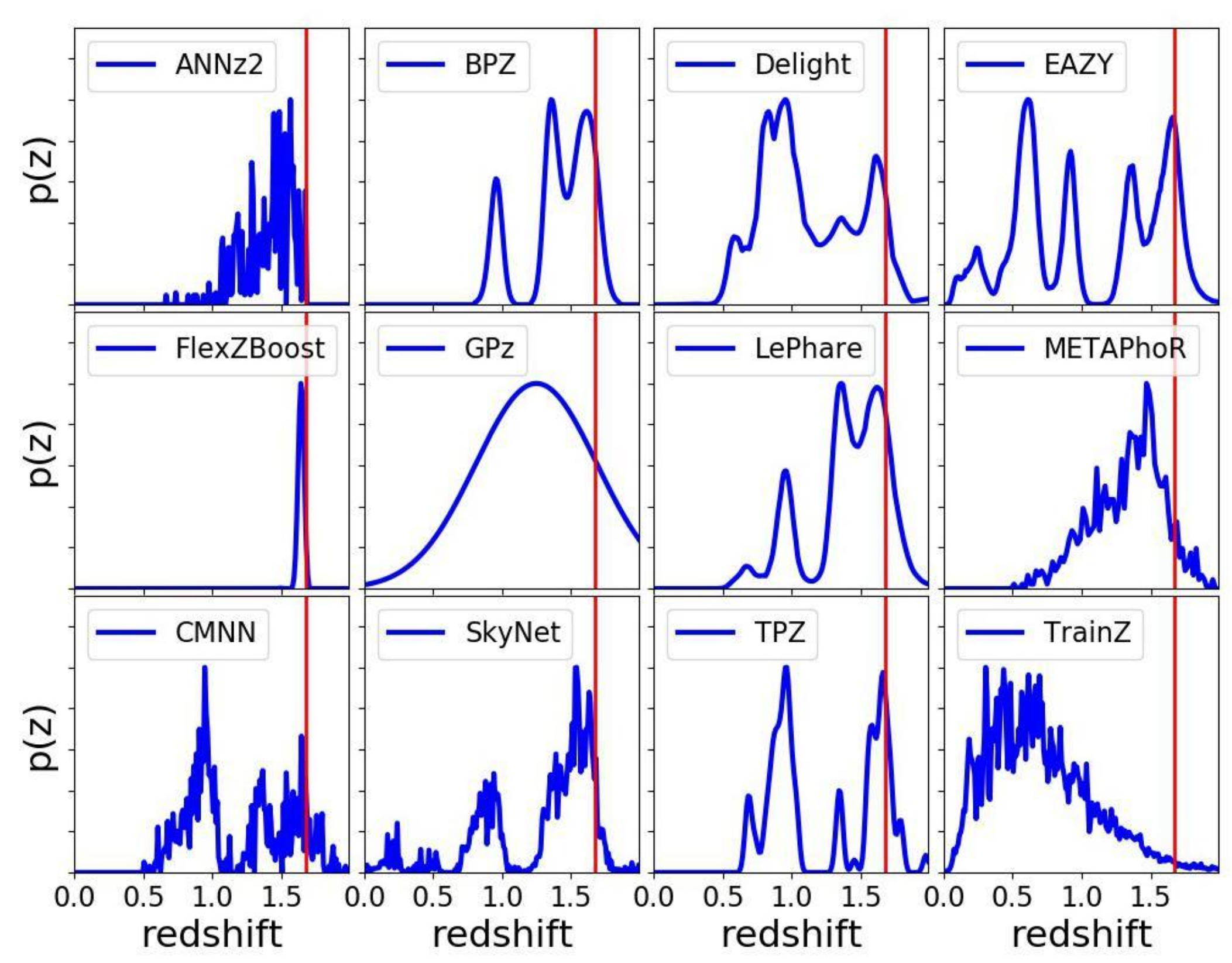}\includegraphics[width=0.49\textwidth]{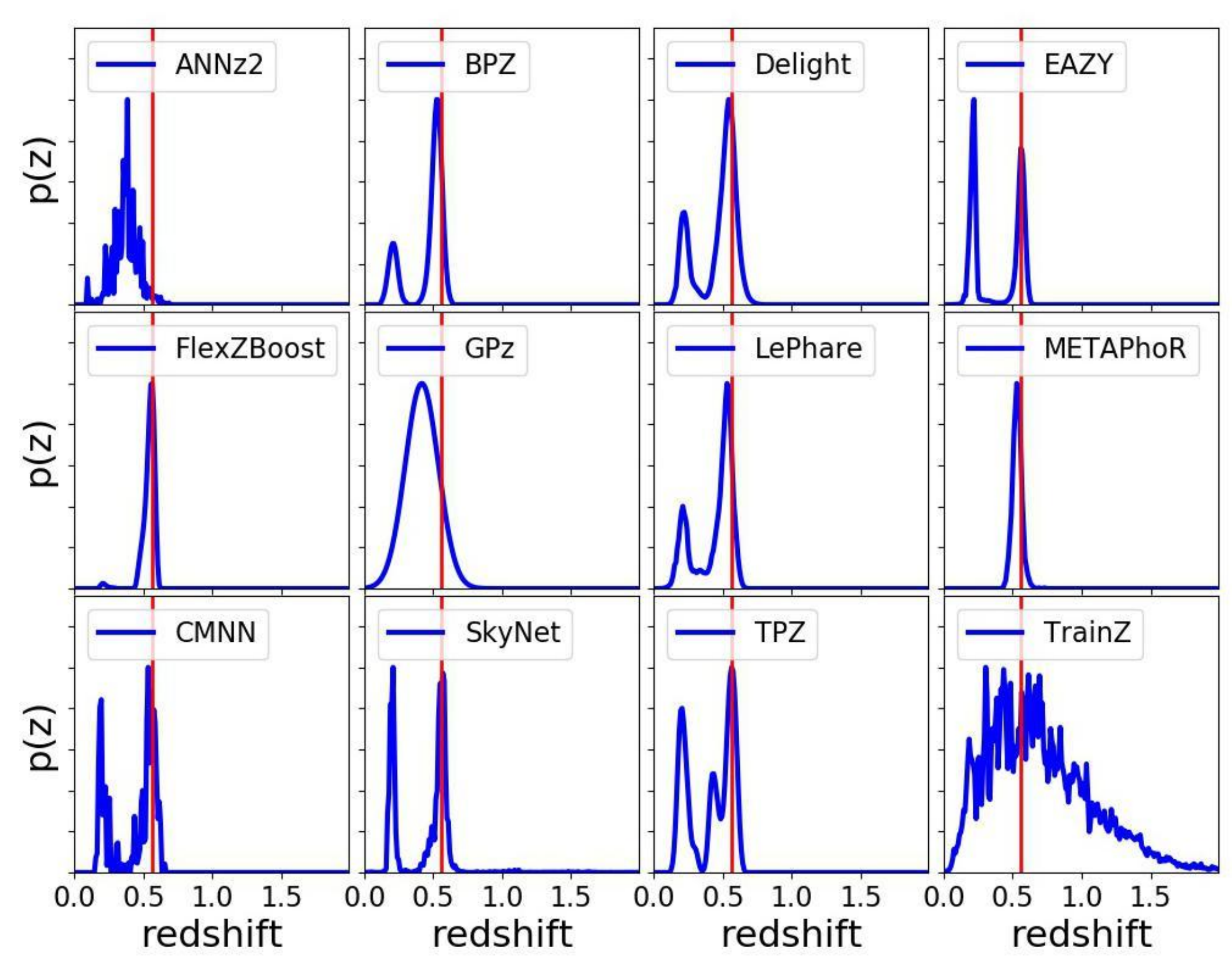}
\caption{The individual \pzpdf s (blue) distributions produced by the twelve codes (small panels) on four exemplary galaxies' photometry (large panels) with different true redshifts (red).
All \pzpdf s have been scaled to the same peak value.
The \pzpdf s of all codes share some features for the example galaxies due to physical colour degeneracies and photometric errors: tight unimodal $p(z)$ (upper left), broad unimodal $p(z)$ (upper right), bimodal $p(z)$ (lower right), and complex/multimodal $p(z)$ (lower left).
The diverse algorithms and implementations induce differences in small-scale structure and sensitivity to physical systematics.}
\label{fig:pz_examples}
\end{figure*}

We begin with a demonstrative visual inspection of the \pzpdf s produced by each code for individual galaxies.
Figure~\ref{fig:pz_examples} shows the \pzpdf s for four galaxies chosen as examples of \pzpdf\ archetypes: a narrow unimodal PDF, a broad unimodal PDF, a bimodal PDF, and a multimodal PDF.
We reiterate that under our idealized experimental conditions, differences between the resulting \pzpdf s are the isolated signature of the implicit prior due to the method by which the \pzpdf s were derived.

The most striking differences between codes are the small-scale features induced by the interaction between the shared piecewise constant parameterization of $200$ bins for $0 < z < 2$ of Section~\ref{sec:metrics} and the smoothing conditions or lack thereof in each algorithm.
The $\mathrm{d}z = 0.01$ redshift resolution is sufficient to capture the broad peaks of faint galaxies' \pzpdf s with large photometric errors but is too broad to resolve the narrow peaks for bright galaxies' \pzpdf s with small photometric errors.
This observation is consistent with the findings of \citet[]{Malz:2018} that the piecewise constant form underperforms other parameterizations in the presence of small-scale structures.

However, the shared small-scale features of \annz, \metaphor, \cmnn, and \skynet\ are a result of various weighted sums of the limited number of training set galaxies with colours similar to those of the test set galaxy in question, with behavior closer to classification than regression in the case of \annz.
The settings used on \gpz\ in this work forced broadening of the single Gaussian to cover the multimodal redshift solutions of the other codes.

\subsection{Performance on \pzpdf\ ensembles}
\label{sec:pitqq}

\begin{figure*}
\centering
\includegraphics[width=0.74\textwidth]{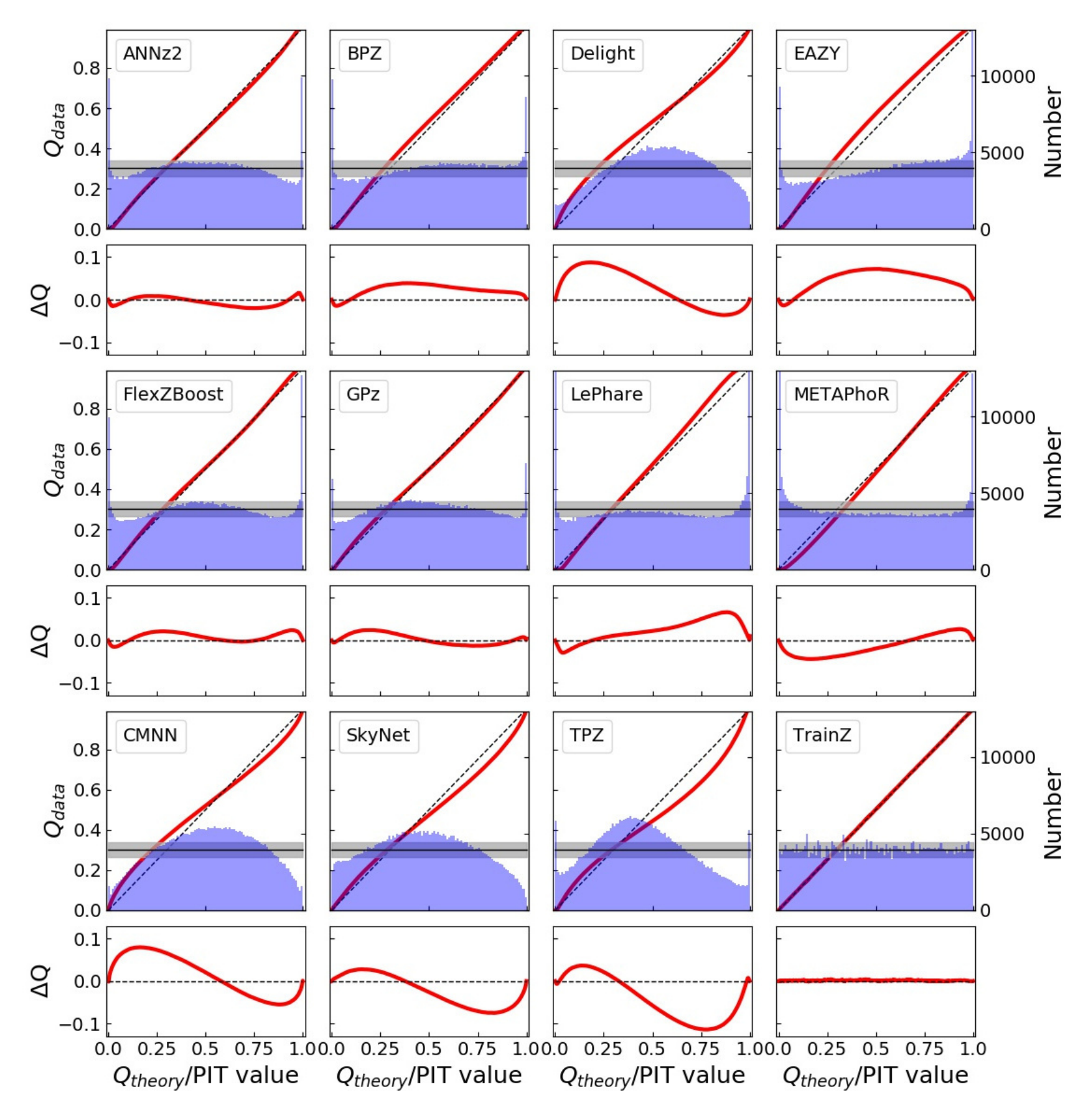}
\caption{The QQ plot (red) and PIT histogram (blue) of the \pzpdf\ codes (panels) along with the ideal QQ (black dashed diagonal) and ideal PIT (gray horizontal) curves, as well as a difference plot for the QQ difference from the ideal diagonal (lower inset).
The gray shaded region indicates the $2 \sigma$ range from a bootstrap resampling of the training set with a size of \xsout{30}\boldblue{44},000 galaxies using \trainz.
The twelve codes exhibit varying degrees of four deviations from perfection: an overabundance of PIT values at the centre of the distribution indicate a catalogue of overly broad \pzpdf s, an excess of PIT values at the extrema indicates a catalogue of overly narrow \pzpdf s, catastrophic outliers manifest as overabundances at PIT values of 0 and 1, and asymmetry indicates systematic bias, a form of model misspecification.
Values in excess of the $2\sigma$ shaded region show that for some codes these errors will be significant given expected training sample sizes.}
\label{fig:pitqq}
\end{figure*}

The histogram of PIT values, QQ plot, and QQ difference plot relative to the ideal diagonal are provided in Figure~\ref{fig:pitqq}, showcasing the biases and trends in the average accuracy of the \pzpdf s for each code.
The high QQ values (i.~e.~ more high than low PIT values) of \bpz, \cmnn, \delight, \eazy, and \gpz\ indicate \pzpdf s biased low, and the low QQ values (more low than high PIT values) of \skynet\ and \tpz\ indicate \pzpdf s biased high.  \boldblue{All three template-based codes use a parameterized functional form for the prior, and deviations from this form in the test set could account for some of the biases seen.  In addition, the \eazy\ and \lephare\ codes both seem to contain an error in how the marginalize over likelihoods, which likely contributes to their observed biases.  For machine learning codes, \delight\ employed a uniform prior in both apparent magnitude and SED type, which could have biased the resultant PDFs, and \gpz\ parameterized all PDFs as a single Gaussian, which may not have been a flexible enough parameterization to capture the full complexity of the posteriors.  In other cases, an obvious source for bias was not discovered, though we note in the particular instance of \tpz\,the results showed a tradeoff in bias versus smoothing width during validation, and a smoothing value that balanced the two effects was chosen.}

The gray shaded band marks the $2\sigma$ variance in PIT values found using the \trainz\ algorithm with a bootstrap resampling of the training set and a sample size of \xsout{30}\boldblue{44},000 galaxies, representing a \xsout{very}conservative estimate of the \boldblue{expected errors due to the training sample size}\xsout{representative training sample size estimated as being required for direct \pz\ calibration  }, and thus an approximate minimal error significance compared to ideal performance.
The existence of deviations in the PIT histograms outside of this gray shaded uncertainty range show that significant biases are present for some codes.

The PIT histograms of \delight, \cmnn, \skynet, and \tpz\ feature an underrepresentation of extreme values, indicative of overly broad \pzpdf s, while the overrepresentation of extreme values for \metaphor\ indicates overly narrow \pzpdf s.
These five codes in particular have a free parameter for bandwidth, which may be responsible for this vulnerability, in spite of the opportunity for fine-tuning with perfect prior information.  \boldblue{In many cases this bandwidth parameter can be thought of as tuning the size of the neighbourhood of influence in the parameter space of the training data.  Choosing too small of a bandwidth can lead to overfitting and overconfidence, leading to narrow PDFs, while choosing too large of a neighbourhood can oversmooth and lead to overly broad PDFs.} 
\flexzboost's ``sharpening'' parameter (described in Section~\ref{sec:flexzboost}) played a key role in diagonalizing the QQ plot, indicating a common avenue for improvement in the approaches that share this type of parameter.
On the other hand, the three purely template-based codes, \bpz, \eazy, and \lephare, do not exhibit much systematic broadening or narrowing, which may indicate that complete template coverage effectively defends from these effects.

\begin{table}
\setlength{\tabcolsep}{2pt}
\centering
\caption{The catastrophic outlier rate as defined by extreme PIT values.
We expect a value of 0.0002 for a proper Uniform distribution.
An excess over this small value indicates true redshifts that fall outside the non-zero support of the $p(z)$.}
\label{tab:pitoutlier}
\begin{tabular}{lc}
\hline
\hline
\Pz\ Code & fraction PIT$<10^{-4}$ or $>$0.9999\\
\hline
\annz       & 0.0265\\
\bpz        & 0.0192\\
\delight    & 0.0006\\
\eazy       & 0.0154\\
\flexzboost & 0.0202\\
\gpz        & 0.0058\\
\lephare    & 0.0486\\
\metaphor   & 0.0229\\
\cmnn       & 0.0034\\
\skynet     & 0.0001\\
\tpz        & 0.0130\\
\hline
\trainz     & 0.0002\\
\end{tabular}
\end{table}

Close inspection of the extremes at PIT values of 0 and 1 reveal spikes in the first and last bin of the PIT histogram for some codes in Figure~\ref{fig:pitqq}, corresponding to catastrophic outliers where the true redshift lies outside of the support of the $p(z)$.
The catastrophic outlier rates are provided in Table~\ref{tab:pitoutlier}.
As expected, \trainz\ achieves precisely the 0.0002 value expected of an ideal PIT distribution.
\annz, \flexzboost, \lephare, and \metaphor\ have notably high catastrophic outlier rates $> 0.02$, exceeding 100 times the ideal PIT rate, meriting further investigation elsewhere.

\begin{figure*}
\centering
\includegraphics[width=0.74\textwidth]{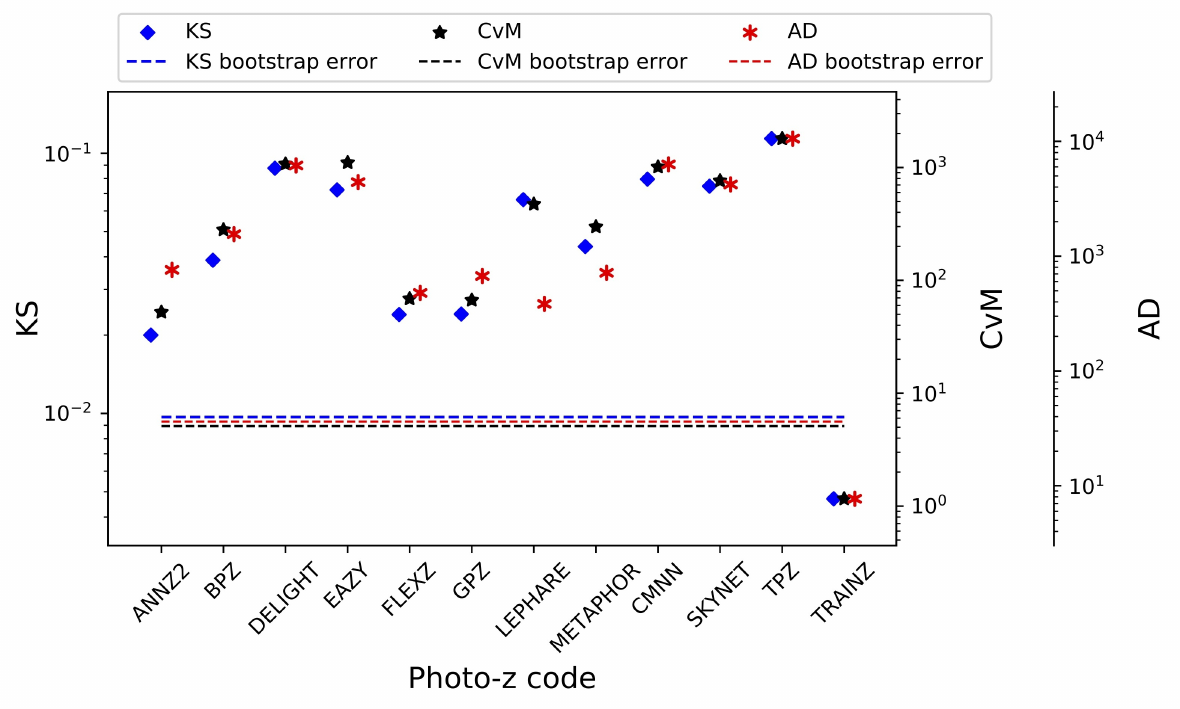}
\caption{A visualization of the Kolmogorov-Smirnoff (KS, blue diamond), Cramer-von Mises (CvM, black star), and Anderson-Darling (AD, red asterisk) statistics for the PIT distributions.
There is generally good agreement between these statistics, with differences corresponding to the codes with outstanding catastrophic outlier rates, a reflection of the differences in how each statistic weights the tails of the distribution.
Horizontal lines indicate the level of uncertainty found by bootstrapping a training set sample of \xsout{30}\boldblue{44},000 galaxies using \trainz; none of the codes reach this conservative ideal floor in expected uncertainty.}
\label{fig:pit_stats}
\end{figure*}

Figure~\ref{fig:pit_stats} highlights the relative values of the KS, CvM, and AD test statistics calculated by comparing the PIT distribution and a uniform distribution $U(0, 1)$.
\metaphor\ and \lephare\ perform well under the AD but poorly under the KS and CvM due to their high catastrophic outlier rates.
\annz\ and \flexzboost\  are the top scorers under these metrics of the PIT distribution.
\annz's strong performance can be attributed to an aspect of the training process in which training set galaxies with PIT values that more closely match the percentiles of the DC1 training set's redshift distribution are upweighted; in effect, these quantile-based metrics were part of the algorithm itself that may or may not serve it well under more realistic experimental conditions.
Similar to what was done for the PIT histograms in Figure~\ref{fig:pitqq}, we create bootstrap training samples of \xsout{30}\boldblue{44},000 galaxies for use with \trainz\ in order to estimate a conservative statistical floor that we would expect in real data.
No code reaches this idealized floor, indicating that all codes suffer some degradation from the ideal when employing their implicit priors, though \annz, \flexzboost, and \gpz\ are within \boldblue{roughly} a factor of two.

\subsection{Performance on individual \pzpdf s}
\label{sec:cdelossresults}

The values of the CDE loss statistic of individual \pzpdf\ accuracy are provided in Table~\ref{tab:cdeloss}.
\boldblue{Even in the absence of knowledge of the true underlying distribution the CDE loss can be determined up to a constant, thus the relative values provide a quantitative measure of code performance.}
It is worth noting that strong performance on the CDE loss, corresponding to lower values of the metric, should imply strong performance on the other metrics, though the inverse is not necessarily true.
Thus the CDE loss is the most effective metric for generic science cases.

\begin{table}  %%% DATA TABLE %%%
\centering
\caption{CDE loss statistic of the individual \pzpdf s for each code.
A lower value of the CDE loss indicates more accurate individual \pzpdf s, with \cmnn\ and \flexzboost\ performing best under this metric.}
\label{tab:cdeloss}
\begin{tabular}{lr}
\hline
Photo-$z$ Code & CDE Loss \\
\hline
\annz     & $-6.88$ \\
\bpz     & $-7.82$ \\
\delight    & $-8.33$\\
\eazy       & $-7.07$ \\
\flexzboost & $-10.60$\\
\gpz    & $-9.93$ \\
\lephare   & $-1.66$ \\
\metaphor & $-6.28$ \\
\cmnn       & $-10.43$ \\
\skynet   & $-7.89$ \\
\tpz     & $-9.55$ \\
\hline
\trainz  & $-0.83$ \\
\end{tabular}
\end{table}

Of the metrics we were able to consider in this experiment, \textit{the CDE Loss is the only metric that can appropriately penalize the pathological \trainz}.
Additionally, it favors \cmnn\ and \flexzboost, the latter of which is optimized for this metric.

\section{Discussion and future work}
\label{sec:discussion}
In contrast with other \pzpdf\ comparison papers that have aimed to identify the ``best'' code for a given survey, we have focused on the somewhat more philosophical questions of how to assess \pzpdf\ methods and how to interpret differences between codes in terms of \pzpdf\ performance.
In Section~\ref{sec:caution}, we reframe the strong performance of our pathological \pzpdf\ technique, \trainz, as a cautionary tale about the importance of choosing appropriate comparison metrics.
In Section~\ref{sec:futureexperiments}, we outline the experiments we intend to build upon this study.
In Section~\ref{sec:futuredata}, we discuss the enhancements of the mock data set that will be necessary to enable the future experiments.

\subsection{Interpretation of metrics}
\label{sec:caution}

We remind the reader that codes utilized in this study were given a goal of obtaining accurate \pzpdf s, not an accurate stacked estimator of the redshift distribution, so we do not expect the same codes to necessarily perform well for both classes of metrics.
Indeed, the codes were optimized for their interpretation of our request for ``accurate \pzpdf s,'' and we expect that the implementations would have been adjusted had we requested optimization of the traditional metrics of Appendices~\ref{sec:moments} and~\ref{sec:pointmetrics}.

Furthermore, our metrics are not necessarily able to assess the fidelity of individual \pzpdf s relative to true posteriors: in the absence of a ``true PDF'' from which redshifts are drawn, it is difficult to construct metrics to measure performance for individual galaxies rather than ensembles.
(The CDE Loss metric of section~\ref{sec:CDE_loss} is an exception to this rule.)
A lack of appropriate metrics more sophisticated than the CDE Loss remains an open issue for science cases requiring accurate individual galaxy PDFs.
The metric-specific performance demonstrated in this paper implies that we may need multiple \pzpdf\ approaches tuned to each metric in order to maximize returns over all science cases in large upcoming surveys.

The \trainz\ estimator of Section~\ref{sec:trainz}, which assigns every galaxy a \pzpdf\ equal to $N(z)$ of the training set, is introduced as an experimental control or null test to demonstrate this point via \textit{reductio ad absurdum}.
Because our training set is perfectly representative of the test set, $N(z)$ should be identical for both sets down to statistical noise.
\textit{We make the alarming observation that \trainz, the absurd experimental control, outperforms all codes on the CDF-based metrics, and all but one code on the $N(z)$ based statistics.}
The PIT and other CDF-based metrics upon which modern \pzpdf\ comparisons are built \citep{Bordoloi:10,Polsterer:16,Tanaka:17} can be gamed by a trivial estimator that yields only an affirmation of prior knowledge uninformed by the data.
\boldblue{The extreme case of \trainz\ also exemplifies the fact that characteristics of the training data are imprinted on the ensemble distributions via the prior, and any imperfections and biases in the training sample will be propagated to the measured \pzpdf s.}
In other words, such ensemble metrics are insufficient for the task of selecting \pzpdf\ codes for analysis pipelines\footnote{That being said, we note that close relatives of \trainz\ have been employed by weak lensing surveys of the past and present \citep{Lima:08, Hildebrandt:17, Hoyle:18} to estimate $N(z)$ by assigning each test set galaxy the redshift distribution of a \emph{subset} of the training set, where the subset is defined as similar to the test set galaxy in the space of photometric data.
The specific science goal naturally guides the choice of metric to focus on $N(z)$ rather than individual \pzpdf s, and on the basis of that metric such improvements upon \trainz\ are guaranteed to be more robust to training set imbalance.}.

The CDE loss and point estimate metrics appropriately penalize \trainz's naivete.
As shown in Appendix~\ref{sec:pointmetrics}, \trainz ~has identical $ZPEAK$ and $ZWEIGHT$ values for every galaxy, and thus the \pz\ point estimates are constant as a function of true redshift, i.e.~a horizontal line at the mode and mean of the training set distribution respectively.
The explicit dependence on the individual posteriors in the calculation of the CDE loss, described in Section~\ref{sec:cdelossresults}, distinguishes this metric from those of the \pzpdf\ ensemble and stacked estimator of the redshift distribution, despite their prevalence in the \pz\ literature.

In summary, context is crucial to defend against deceptively strong performers such as \trainz;
the best \pzpdf\ method is the one that most effectively achieves the science goals of a particular study, not the one that performs best on a metric that does not reflect those goals.
In the absence of a single scientific motivation or the information necessary for a principled metric definition, we must consider many metrics and be critical of the information transmitted by each.

\subsection{Extensions to the experimental design}
\label{sec:futureexperiments}

The work presented in this paper is only a first step in assessing \pzpdf\ approaches and moving toward a photometric redshift estimator that will be employed for LSST analyses.  Extensions of the experimental design will require further rounds of analyses, and the authors welcome interest from those outside \lsstdesc\ to have their codes assessed in these future investigations.

This initial paper explores \pzpdf\ code performance in idealized conditions with perfect catalogue-based photometry and representative training data, but the resilience of each code to realistic imperfections in prior information has not yet been evaluated.
A top priority for a follow-up study is to test realistic forms of incomplete, erroneous, and non-representative template libraries and training sets as well as the impact of other forms of external priors that must be ingested by the codes, major concerns in \citet{Newman:2015, Masters:2017}.
Outright redshift failures due to emission line misidentification or noise spikes may be modeled by the inclusion of a small number of high-confidence yet false redshifts.
We plan to perform a full sensitivity analysis on a realistically incomplete training set of spectroscopic galaxies, modeling the performance of spectrographs, emission-line properties, and expected signal-to-noise to determine which potential training set galaxies are most likely not to yield a secure redshift.

Appendix~\ref{sec:moments} only addresses the stacked estimator of the redshift distribution of the entire galaxy catalogue rather than subsets in bins, tomographic or otherwise.
The effects of tomographic binning schemes will be explored in a dedicated future paper, including propagation of redshift uncertainties in a set of fiducial tomographic redshift bins in order to estimate impact on cosmological parameter estimation.

Sequels to this study will also address some shortcomings of our experimental procedure.
The fixed redshift grid shared between the codes may have unfairly penalized codes with a different native parameterization, as precision is lost when converting between formats.
Performance on the (admittedly small) population of sharply peaked \pzpdf s may have been suppressed across all codes due to the insufficient resolution of the redshift grid.
In light of the results of \citet[]{Malz:2018}, in future analyses we plan to switch from a fixed grid to the quantile parameterization or to permit each code to use its native storage format under a shared number of parameters.

Section~\ref{sec:metrics} discussed the difficulty in evaluating PDF accuracy for individual objects with known $(z, d)$ information but without a known $p(z, d)$.
In a follow-up study, we will
generate mock data probabilistically, yielding true PDFs in addition to true redshifts and photometric data.
This future data set will enable tests of PDF accuracy for individual galaxies rather than solely ensembles.

\subsection{Realistic mock data}
\label{sec:futuredata}

To make optimal use of the \lsst\ data for cosmological and other astrophysical analyses of the \lsstdesc\ SRM, future investigations that build upon this one will require a more sophisticated set of galaxy photometry and redshifts.
This initial paper explored a data set that was constructed at the catalogue level, with no inclusion of the complications that arise from photometric measurements of imaging data.
Future data challenges will move to catalogues constructed from mock images, including the complications of deblending, sensor inefficiencies, and heterogeneous observing conditions, all anticipated to affect the measured colours of \lsst's galaxy sample \citep{Dawson:2016}.

The DC1 galaxy SEDs were linear combinations of just five basis SED templates, and the next generation of data for \pzpdf\ investigations must include a broader range of physical properties.
Though we only considered $z < 2$ here, \lsst\ 10-year data will contain $z > 2$ galaxies, plagued by fainter apparent magnitudes and anomalous colours due to stellar evolution.
A subsequent study must also have a data set that includes low-level active galactic nuclei (AGN) features in the SEDs, which perturb colours and other host galaxy properties.
An observational degeneracy between the Lyman break of a $z \sim 2-3$ galaxy from the Balmer break of a $z \sim 0.2-0.3$ galaxy is a known source of catastrophic outliers \citep{Massarotti:2001} that was not effectively included in this study.
To gauge the sensitivity of \pzpdf\ estimators to catastrophic outliers, our data set must include realistic high-redshift galaxy populations.

\section{Conclusion}
\label{sec:conclusion}

This paper compares twelve \pzpdf\ codes under controlled experimental conditions of representative and complete prior information to set a baseline for an upcoming sensitivity analysis.
This work isolates the impact on metrics of \pzpdf\ accuracy due to the estimation technique as opposed to the complications of realistic physical systematics of the photometry.
Though the mock data set of this investigation did not include true \pz\ posteriors for comparison, \textit{we interpret deviations from perfect results given perfect prior information as the imprint of the implicit assumptions underlying the estimation approach}.

We evaluate the twelve codes under science-agnostic metrics both established and emerging to stress-test the ensemble properties of \pzpdf\ catalogues derived by each method.
In appendices, we also present metrics of point estimates and a prevalent summary statistic of \pzpdf\ catalogues used in cosmological analyses to enable the reader to relate this work to studies of similar scope.
We observe that no one code dominates in all metrics, and that the standard metrics of \pzpdf s and the stacked estimator of the redshift distribution can be gamed by a very simplistic procedure that asserts the prior over the data.
We emphasize to the \pz\ community that \textit{metrics used to vet \pzpdf\ methods must be scrutinized to ensure they correspond to the quantities that matter to our science}.

% ----------------------------------------------------------------------

\subsection*{Acknowledgments}

%%% Here is where you should add your specific acknowledgments, remembering that some standard thanks will be added via the \code{desc-tex/ack/*.tex} and \code{contributions.tex} files.
The authors would like to thank the journal referee, who made many insightful suggestions that improved the draft.
This paper has undergone internal review in the LSST Dark Energy Science Collaboration.
The authors acknowledge feedback from the internal reviewers: Daniel Gruen, Markus Rau, and Michael Troxel. % REQUIRED if true

Author contributions are listed below. \\
S.J.~Schmidt: Co-led the project. (conceptualization, data curation, formal analysis, investigation, methodology, project administration, resources, software, supervision, visualization, writing -- original draft, writing -- review \& editing). \\
A.I.~Malz: Co-led the project, contributed to choice of metrics, implementation in code, and writing. (conceptualization, methodology, project administration, resources, software, visualization, writing -- original draft, writing -- review \& editing). \\
J.Y.H.~Soo: Ran ANNz2 and Delight, updated abstract, edited sections 1 through 6, added tables in Methods and Results, updated references.bib and added references throughout the paper. \\
I.A.~Almosallam: vetted the early versions of the data set and ran many photo-z codes on it, applied GPz to the final version and wrote the GPz subsection. \\
M.~Brescia: Main-ideator of MLPQNA and co-ideator of METAPHOR; modification of METAPHOR pipeline to fit the LSST data structure and requirements. \\
S.~Cavuoti: Co-ideator of METAPHOR, contributed to choice and test of metrics, ran METAPHOR, minor text editing. \\
J.~Cohen-Tanugi: contributed to running code, analysis discussion, and editing, reviewing the paper. \\
A.J.~Connolly: Developed the colour-matched nearest-neighbours photo-z code; participated in discussions of the analysis. \\
J.~DeRose: One of the primary developers of the Buzzard-highres simulated galaxy catalogue employed in the analysis. \\
P.E.~Freeman: Contributed to choice of CDE metrics and to implementation of FlexZBoost. \\
M.L.~Graham: Ran the colour-matched nearest-neighbours photo-z code on the Buzzard catalogue and wrote the relevant piece of Section 2; participated in discussions of the analysis. \\
K.G.~Iyer: assisted in writing metric functions used to evaluate codes. \\
M.J.~Jarvis: Contributed text on AGN to Discussion section and portions of GPz work. \\
J.B.~Kalmbach: Worked on preparing the figures for the paper. \\
E.~Kovacs: Ran simulations, discussed data format and properties for SEDs, dust, and ELG corrrections. \\
A.B.~Lee: Co-developed FlexZBoost and the CDE loss statistic, wrote text on the work, and supervised the development of FlexZBoost software packages. \\
G.~Longo: Scientific advise, test and validation of the modified METAPHOR pipeline, text of the METAPHOR section. \\
C.B.~Morrison: Managerial support; Discussions with authors regarding metrics and style; Some coding contribution to metric computation. \\
J.A.~Newman: Contributions to overall strategy, design of metrics, and supervision of work done by Rongpu Zhou. \\
E.~Nourbakhsh: Ran and optimized TPZ code and wrote a subsection of Section 2 for TPZ. \\
E.~Nuss: contributed to running code, analysis discussion, and editing,reviewing the paper. \\
T.~Pospisil: Co-developed FlexZBoost software and CDE loss calculation code. \\
H.~Tranin: contributed to providing SkyNet results and writing the relevant section. \\
R.H.~Wechsler: Project lead for Buzzard-highres simulated galaxy catalogue employed in analysis. \\
R.~Zhou: Optimized and ran EAZY and contributed to the draft. \\
R.~Izbicki: Co-developed FlexZBoost and the CDE loss statistic, and wrote software for FlexZBoost \\

% This work used TBD kindly provided by Not-A-DESC Member and benefitted from comments by Another Non-DESC person.

% Standard papers only: A.B.C. acknowledges support from grant 1234 from ...

The authors express immense gratitude to Alex Abate, without whom this paper would not have gotten started.  We would also like to thank Ofer Lahav and Will Hartley for contributions during the writing of the draft.
% The authors would like to thank their LSST-DESC publication review committee for comments that improved the paper draft.
% hack event acknowledgments are important
We thank Stony Brook University for hosting the Summer 2017 LSST-DESC Hack Day at which this work was partially completed.

SJS and EN acknowledge support from DOE grant DE-SC0009999.  SJS acknowledges support from NSF/AURA grant N56981C.
AIM acknowledges support from the Max Planck Society and the Alexander von Humboldt Foundation in the framework of the Max Planck-Humboldt Research Award endowed by the Federal Ministry of Education and Research.
During the completion of this work, AIM was advised by David W. Hogg and was supported by National Science Foundation grant AST-1517237.
JYHS acknowledges financial support from the MyBrainSc Scholarship (Ministry of Education, Malaysia), and the supervision of Ofer Lahav and Benjamin Joachimi.
JYHS would also like to thank Antonella Palmese and Boris Leistedt for guidance on the use of the algorithms ANNz2 and Delight respectively.
I.A. acknowledges the support of King Abdulaziz City for Science and Technology.
MB acknowledges support from the Agreement ASI/INAF 2018-23-HH.0 - Phase D.
AJC and JBK acknowledges support from DOE grant DE-SC-0011635. AJC, JBK, MLG, CBM acknowledge support from the DIRAC Institute in the Department of Astronomy at the University of Washington. The DIRAC Institute is supported through generous gifts from the Charles and Lisa Simonyi Fund for Arts and Sciences, and the Washington Research Foundation.
PEF acknowledges support from NSF grant 1521786.
MJJ acknowledges support from Oxford Hintze Centre for Astrophysical Surveys which is funded through generous support from the Hintze Family Charitable Foundation.
ABL and TP acknowledge support from NSF DMS grant 1520786.
GL acknowledges partial funding from the EU funded ITN Marie Curie Network SUNDIAL.
CBM is supported in part by the National Science Foundation through Cooperative Agreement 1258333 managed by the Association of Universities for Research in Astronomy(AURA), and the Department of Energy under Contract No. DE-AC02-76SF00515 with the SLAC National Accelerator Laboratory. Additional LSST funding comes from private donations, grants to universities, and in-kind support from LSSTC Institutional Members.
JAN and RZ were supported by the U.S. Department of Energy, Office of Science, Office of High Energy Physics under award number DE-SC0007914.
RI acknowledges support from FAPESP grant 2019/11321-9 and CNPq grant 306943/2017-4.

In addition to packages cited in the text, analyses performed in this paper used the following software packages: \code{Numpy} and \code{Scipy} \citep{numpyscipy}, \code{Matplotlib} \citep{matplotlib}, \code{Seaborn} \citep{seaborn}, \code{minFunc} \citep{minfunc}, \code{qp} \citep{Malz:qp, Malz:2018}, \code{pySkyNet} \citep{pyskynet}, and \code{photUtils} from the \lsst\ simulations package \citep{lsstphotutils}.

The DESC acknowledges ongoing support from the Institut National de Physique Nucl\'eaire et de Physique des Particules in France; the Science \& Technology Facilities Council in the United Kingdom; and the Department of Energy, the National Science Foundation, and the LSST Corporation in the United States.  DESC uses resources of the IN2P3 Computing Center (CC-IN2P3--Lyon/Villeurbanne - France) funded by the Centre National de la Recherche Scientifique; the National Energy Research Scientific Computing Center, a DOE Office of Science User Facility supported by the Office of Science of the U.S.\ Department of Energy under Contract No.\ DE-AC02-05CH11231; STFC DiRAC HPC Facilities, funded by UK BIS National E-infrastructure capital grants; and the UK particle physics grid, supported by the GridPP Collaboration.  This work was performed in part under DOE Contract DE-AC02-76SF00515.

\appendix

\section{Evaluation of the redshift distribution}
\label{sec:moments}

Perhaps the most popular application of \pzpdf s is the estimation of the overall redshift distribution $N(z)$, a quantity that enters some cosmological calculations and the true value of which is known for the DC1 data set and will be denoted as $\tilde{N}(z)$.
In terms of the prior information provided to each method, the true redshift distribution satisfies the tautology $\tilde{N}(z) = p(z \vert I_{D})$ due to our experimental set-up; because the DC1 training and template sets are representative and complete, $I_{D}$ represents a prior that is also equal to the truth.
In this ideal case of complete and representative prior information, the method that would give the best approximation to $\hat{N}(z)$ would be one that neglects all the information contained in the photometry $\{d_{i}\}_{N_{tot}}$ and gives every galaxy the same \pzpdf\ $\hat{p}_{i}(z) = \tilde{N}(z)$ for all $i$; the inclusion of any information from the photometry would only introduce noise to the optimal result of returning the prior.
This is the exact estimator, \trainz, that we have described in Section~\ref{sec:trainz}, and which will serve as an experimental control.

\subsection{Metrics of the stacked estimator of the redshift distribution}
\label{sec:stackedmetrics}

``Stacking'' according to
\begin{equation}
  \label{eq:stacked}
  \hat{N}^{H}(z) \equiv \frac{1}{N_{tot}}\ \sum_{i}^{N_{tot}}\ \hat{p}^{H}_{i}(z)
\end{equation}
 is the most widely used method for obtaining $\hat{N}^{H}(z)$ as an estimator of the redshift distribution from \pzpdf s derived by a method $H$. While the stacked estimator of the redshift distribution violates the mathematical definition of statistical independence and is thus not formally correct\footnote{Malz \& Hogg (in prep) shows how the stacking procedure can lead to bias in the estimate of $N(z)$ and presents a principled alternative to this commonly employed method.  See \url{https://github.com/aimalz/chippr} for details.}, we use it as a basis for comparison of \pzpdf\ methods under the untested assumption that the response of our metrics of $\hat{N}^{H}(z)$ will be analogous to the same metrics applied to a principled estimator of the redshift distribution.

As $N(z)$ is itself a univariate PDF, we apply the metrics of the previous sections to it as well.
We additionaly calculate the first three moments
\begin{equation}
  \label{eq:moment}
  \langle z^{m}\rangle \equiv \int_{-\infty}^{\infty} z^{m} N(z) dz
\end{equation}
of the estimated redshift distribution $\hat{N}^{H}(z)$ for each code and compare them to the moments of the true redshift distribution $\tilde{N}(z)$.
Under the assumption that the stacked estimator is unbiased, a superior method minimizes the difference between the true and estimated moments.

\subsection{Performance on the stacked estimator of the redshift distribution}
\label{sec:stackedmetrics_results}

\begin{figure*}
\centering
\includegraphics[width=0.74\textwidth]{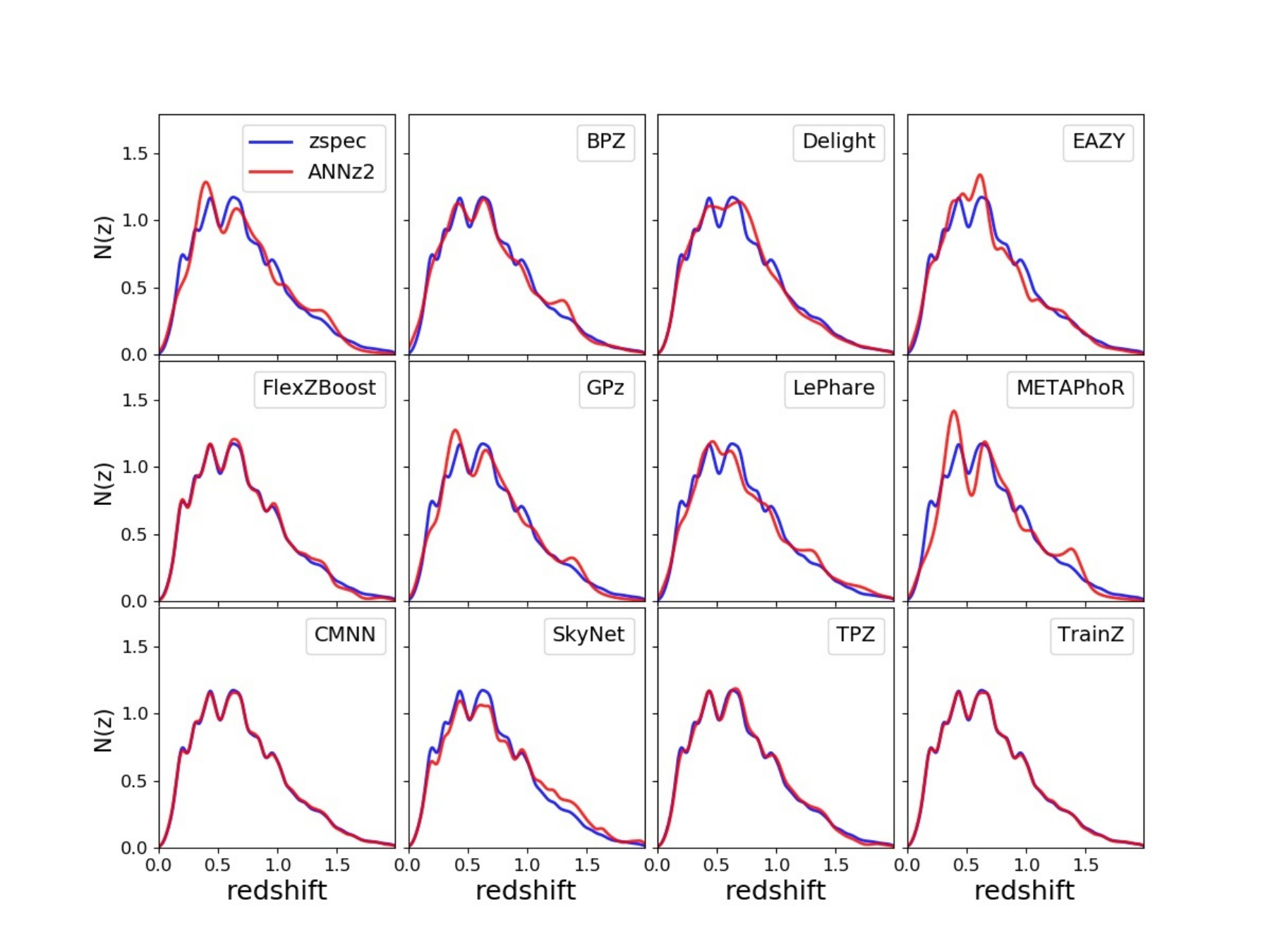}
\caption{The smoothed stacked estimator $\hat{N}(z)$ of the redshift distribution (red) produced by each code (panels) compared to the true redshift distribution $\tilde{N}(z)$ (blue).
Varying levels of agreement are seen among the codes, with the smallest deviations for \cmnn, \flexzboost, \tpz, and \trainz.}
\label{fig:nz}
\end{figure*}

Figure~\ref{fig:nz} shows the stacked estimator $\hat{N}(z)$ of the redshift distribution for each code compared to the true redshift distribution $\tilde{N}(z)$, where the stacked estimator has been smoothed for each code in the plot using a kernel density estimate (KDE) with a bandwidth chosen by Scott's Rule \citep{Scott:1992} in order to minimize visual differences in small-scale features; the quantiative statistics, however, are calculated using the empirical CDF which is not smoothed.

Many of the codes, including all the model-fitting approaches and \annz, \gpz, \metaphor, and \skynet\ from the data-driven camp, overestimate the redshift density at $z \sim 1.4$.
This behavior is a consequence of the $4000$ \AA\ break passing through the gap between the $z$ and $y$ filters, which induces a\xsout{genuine discontinuity}\boldblue{dramatic rise and fall} in the $z - y$ colour as a function of redshift\boldblue{.  With a dearth of strong spectral features blue-ward of the 4000 \AA\ break in most galaxy SEDs, the degeneracies on either side of the peak in $z - y$ colour tend to broaden the redshift PDFs near $z \sim 1.4$, which can lead to the ``bump'' seen in the stacked $\hat{N}(z)$ estimate.}\xsout{that can sway the \pzpdf\ estimates in the absence of bluer spectral features.}

\annz, \gpz, and \metaphor\ feature exaggerated peaks and troughs relative to the training set, a potential sign of overtraining.
Further investigation on overtraining is needed, if present this is an obstacle that may be overcome with adjustment of the implementation.

As expected, \trainz\ perfectly recovers the true redshift distribution: as the training sample is selected from the same underlying distribution as the test set, the redshift distributions are identical, up to Poisson fluctuations due to the finite number of sample galaxies.
\cmnn\ is also in excellent agreement for similar reasons: with a representative training sample of galaxies spanning the colour-space, the sum of the colour-matched neighbour redshifts should return the true redshift distribution.
\flexzboost\ and \tpz\ also perform superb recovery of the true redshift distribution, with only a slight deviation at $z \sim 1.4$.
\xsout{Our metrics, however, cannot discern whether these four approaches, as well as \delight, are spared the $z \sim 1.4$ degeneracy in $\hat{N}(z)$ because they have more effectively used information in the data or if the impact is simply washed out by the stacked estimator's effective average over the test set galaxy sample.
See Appendix~\ref{sec:pointmetrics} for further discussion of the $z \sim 1.4$ issue.}

\begin{figure*}
\centering
\includegraphics[width=0.74\textwidth]{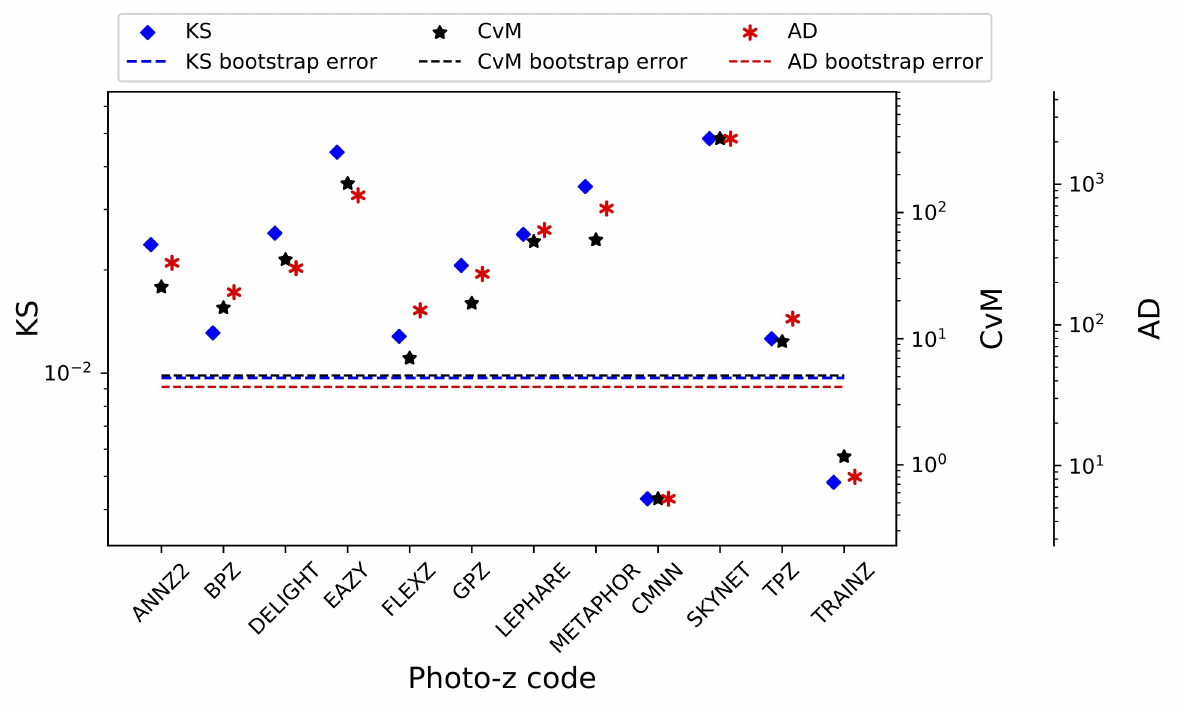}
\caption{A visualization of the Kolmogorov-Smirnoff (KS, blue diamond), Cramer-von Mises (CvM, black star), and Anderson-Darling (AD, red asterisk) statistics for the $\hat{N}(z)$ distributions.
Horizontal lines indicate the statistic values (including uncertainty) achieved using \trainz\ via bootstrap resampling a training set containing\xsout{30}\boldblue{44},000 redshifts.
We make the reassuring observation that these related statistics do not disagree significantly with one another.
\cmnn\ outperforms the control case, \trainz, and several codes are within a factor of two of this conservative idealized limit.
\skynet\ scores poorly due to an overall bias in its redshift predictions.}
\label{fig:nz_stats}
\end{figure*}

Figure~\ref{fig:nz_stats} shows the quantitative Kolmogorov-Smirnoff (KS), Cramer-Von Mises (CvM), and Anderson Darling (AD) test statistics for each of the codes for the $\hat{N}(z)$ based measures.
The horizontal lines show the the result of a bootstrap resampling of the training set using\xsout{30}\boldblue{44},000 samples for \trainz, representing a conservative\xsout{idealized limit on expected performance for a}\boldblue{error estimate assuming our} modest-sized representative training set of galaxies, as mentioned in Section~\ref{sec:pitqq}.
The AD bootstrap statistic is elevated due to its sensitivity to the tails of distributions.
The stacked estimators of the redshift distribution for \cmnn\ and \trainz\ best estimate $\tilde{N}(z)$ under these metrics, whereas \eazy, \lephare, \metaphor, and \skynet\ underperform; \bpz, \gpz, and \tpz\ are within a factor of two of the conservative limit for all statistics.
It is unsurprising that \cmnn\ scores well, as with a nearly complete and representative training set choosing neighbouring points in colour/magnitude space to construct an estimator should lead to excellent agreement in the final $\hat{N}(z)$.

It is, however, surprising that \tpz\ does well on $\hat{N}(z)$ given its poor performance on the ensemble \pzpdf s, especially knowing that \tpz\ was optimized for \pzpdf\ ensemble metrics rather than the stacked estimator of the redshift distribution.
A possible explanation is the choice of smoothing parameter chosen during validation, which affects \pzpdf\ widths as well as overall redshift bias and could be modified to improve performance under the \pzpdf\ metrics.

\begin{figure*}
\centering
\includegraphics[width=0.75\textwidth]{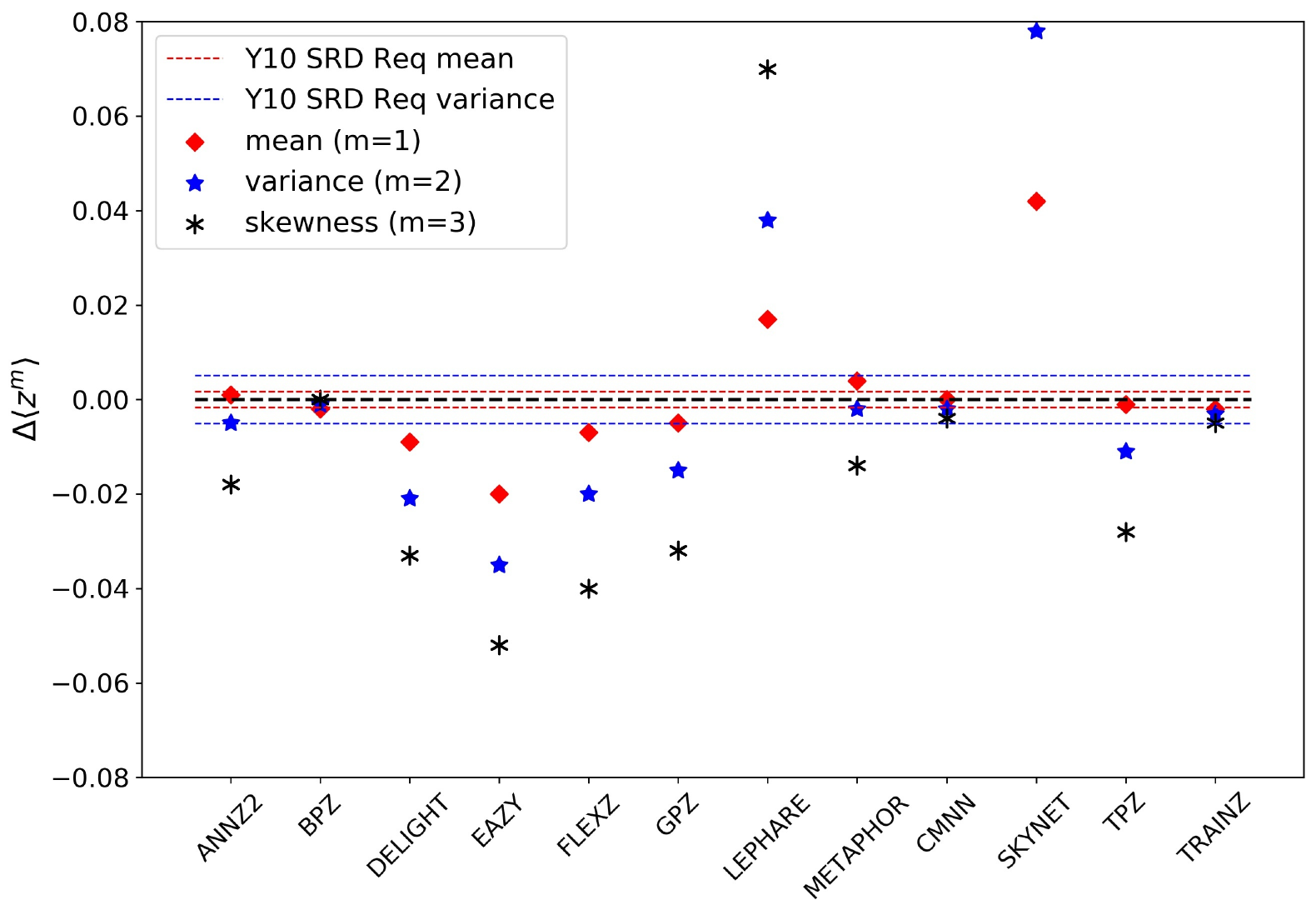}
\caption{Residuals of the first three moments of the stacked $\hat{N}(z)$ distribution.  Red and blue horizontal lines indicate the Year 10 DESC SRD requirements on accuracy of the mean and variance respectively.  Only a small number of codes are able to meet these specifications even with perfect training data.   }
\label{fig:moments}
\end{figure*}

We calculated the first three moments of the stacked $\hat{N}(z)$ distribution of all galaxies and compared it to the moments of the true redshift distribution.  Figure~\ref{fig:moments} shows the residuals of the moments for all codes.  Accuracy of the moments varies widely between codes, raising concerns about the propagation to cosmological analyses.  The DESC SRD \citep{Mandelbaum:2018} lists stringent requirements on how well the mean and variance of tomographic redshift bins must be known for each of the main DESC science cases.  We indicate the Year 10 (Y10) requirements assuming our true mean redshift of $z=0.701$ as dashed lines.  In this study with representative training data, \annz, \cmnn, \tpz, and our pathological \trainz\ estimator meet the Y10 requirement on the mean redshift.  Only \annz, \cmnn, and \trainz\ meet both requirements.  One should be concerned that many codes fail to meet this ambitious limit under perfect prior information because all codes are anticipated to do no better under realistically imperfect prior information, and indicates that additional calibration to remove these systematic offsets \citep[e.~g.~][]{Newman:2008} will likely be necessary in order to meet these stringent goals.

\skynet\ exhibits redshift bias in Figure~\ref{fig:nz} and is a clear outlier in the first moment of $\hat{N}(z)$ in Figure~\ref{fig:moments}.
The \skynet\ algorithm employs a random subsampling of the training set without testing that the subset is representative of the full population, and the implementation used here does not upweight rarer low- and high-redshift galaxies, as in \citet{Bonnett:15}, suggesting a possible cause that may be addressed in future work.

\section{\Pz\ point estimation and metrics}
\label{sec:pointmetrics}

While this work assumes that science applications value the information of the full \pzpdf, we present conventional metrics of \pz\ point estimates as a quick and dirty visual diagnostic tool and to facilitate direct comparisons to historical studies.

\subsection{Reduction of \pzpdf s to point estimates}
\label{sec:pointest}

Though we acknowledge that many of the codes can also return a native \pz\ point estimate, we put all codes on equal footing by considering two generic \pz\ point estimators, the mode $z_{PEAK}$ and main-peak-mean $z_{WEIGHT}$ \citep{Dahlen:13}, a weighted mean within the bounds of the main peak, as identified by the roots of $p(z) - 0.05 \times z_{PEAK}$.
Though $z_{WEIGHT}$ neglects information in a secondary peak of e.~g.~ a bimodal distribution, it avoids the pitfall of reducing the \pzpdf\ to a redshift between peaks where there is low probability.

\subsection{Metrics of \pz\ point estimates}
\label{sec:point_metrics}

We calculate the commonly used point estimate metrics of the overall intrinsic scatter, bias, and catastrophic outlier rate, defined in terms of the standard error $e_{z} \equiv (z_{PEAK} - z_{\mathrm{true}}) / (1 + z_{\mathrm{true}})$.
Because the standard deviation of the \pz\ residuals is sensitive to outliers, we define the scatter in terms of the Interquartile Range (IQR), the difference between the 75th and 25th percentiles of the distribution of $e_{z}$, imposing the scaling $\sigma_{\mathrm{IQR}} = \mathrm{IQR} / 1.349$ to ensure that the area within $\sigma_{\mathrm{IQR}}$ is the same as that within one standard deviation from a standard Normal distribution.
We also resist the effect of catastrophic outliers by defining the bias $b_{z}$ as the median rather than mean value of $e_{z}$.
The catastrophic outlier rate $f_{\mathrm{out}}$ is defined as the fraction of galaxies with $e_{z}$ greater than $\max(3 \sigma_{\mathrm{IQR}}, 0.06)$.

For reference, Section 3.8 of the \lsst\ Science Book \citep{Abell:09} uses the standard definitions of these parameters in requiring
\begin{itemize}
\item RMS scatter $\sigma < 0.02 (1 + z_{\mathrm{true}})$
\item bias $b_{z} < 0.003$ 
\item catastrophic outlier rate $f_{\mathrm{out}} < 10$ per cent 
\end{itemize}

\subsection{Comparison of \pz\ point estimate metrics}
\label{sec:pointmetrics_results}

\begin{figure*}
\centering
\includegraphics[width=0.49\textwidth]{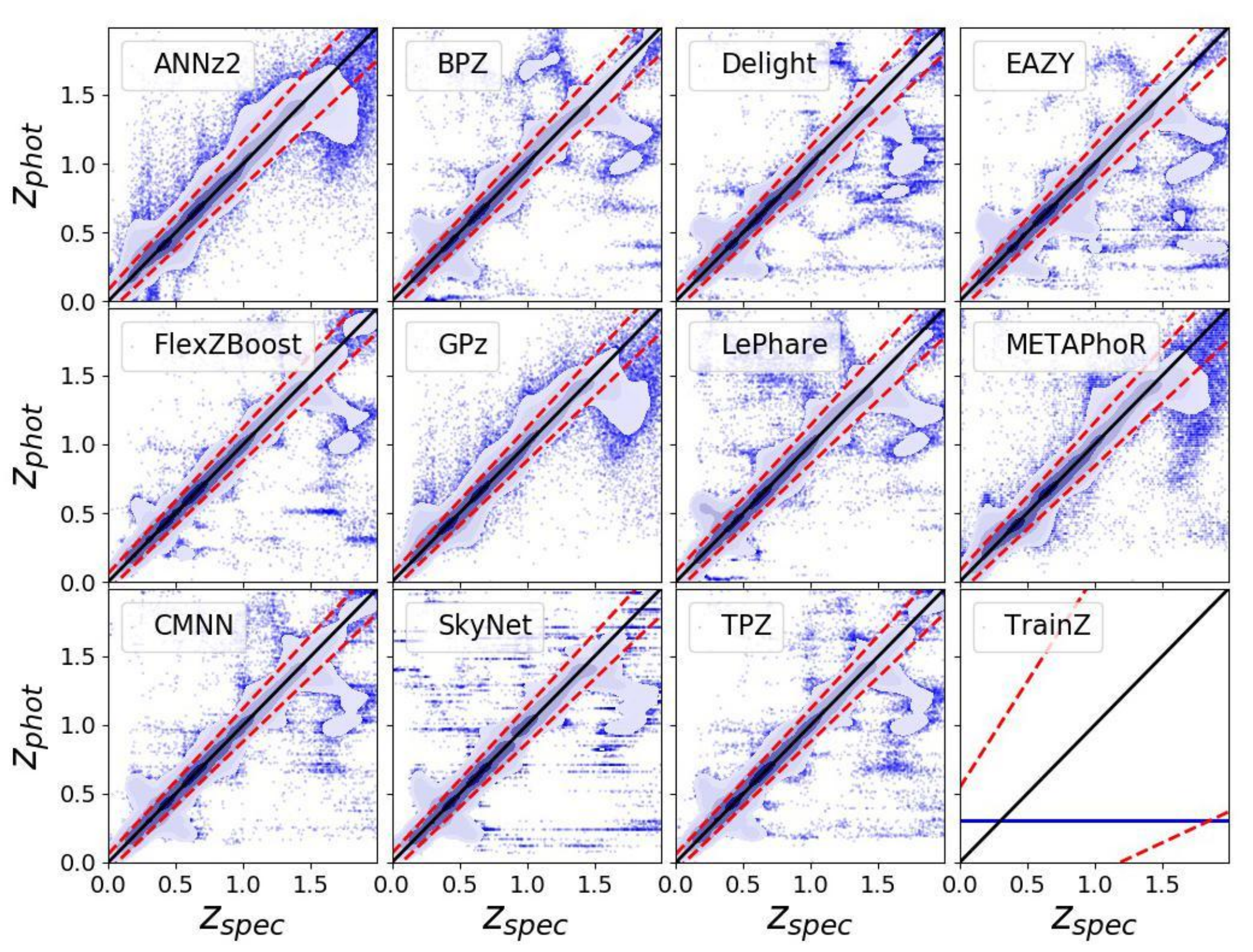}
\includegraphics[width=0.49\textwidth]{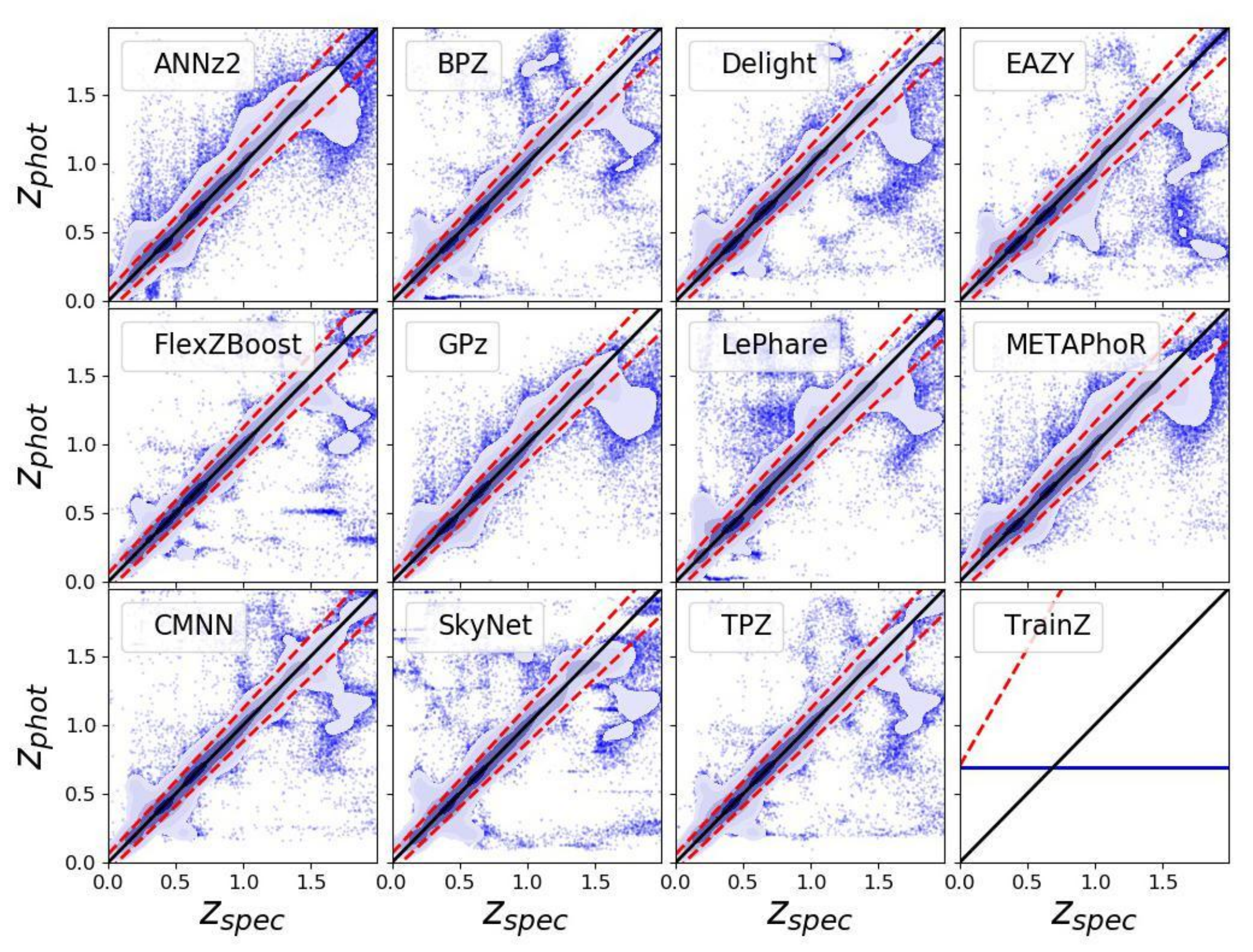}
\caption{The density of \pz\ point estimates (contours) reduced from the \pzpdf s with outliers (blue) beyond the outlier cutoff (red dashed lines), via the mode ($z_{PEAK}$, left panel) and main-peak-mean ($z_{WEIGHT}$, right panel).
The \trainz\ estimator (lower right sub-panels) has a shared $z_{PEAK}$ and $z_{WEIGHT}$ for the entire test set galaxy sample.}
\label{fig:pz_pointestimates}
\end{figure*}

Figure~\ref{fig:pz_pointestimates} shows\xsout{both} point estimates\xsout{both} $z_{PEAK}$ and $z_{WEIGHT}$ \boldblue{versus true redshift for all codes}.
Point density is shown with mixed contours to emphasize that most of the galaxies do fall close to the $z_{phot} = z_{spec}$ line, while points trace the details of the catastrophic outlier populations.

The finite grid spacing of the \pzpdf s induces some discretization in $z_{PEAK}$.
The features perpendicular to the $z_{phot} = z_{spec}$ line are due to the $4000$ \AA\ break passing through the gaps between adjacent filters\boldblue{,  most notably at $z \sim 0.4$ between the g and r filters, and at $z \sim 1.4$ between the z and y filters.  All codes show some level of concentrated catastrophic outliers, often due to degenerate redshift solutions for a single area of color space.  In the presence of genuine ambiguities in the colour space to redshift mapping, the compression of information down to a single number that is the point estimate redshift fails to adequately capture the more complex nature of the multimodal redshift matching: using the mode selects the highest peak while ignoring posterior probability in lower valued redshift peaks, while measuring the mean of a multimodal PDF often results in a value lying at a redshift between the peaks, removed from either/any of the most likely redshift solutions.  Thus,}\xsout{E}even the strongest codes feature populations far from the $z_{phot} = z_{spec}$ line representing a \boldblue{genuine} degeneracy in the space of colours and redshifts.

\begin{table*}
\begin{center}
\caption{\Pz\ point estimate statistics}\label{tab:pointestimates}
\begin{tabular}{lcccccc}
\hline
\hline
                 &            & $Z_{PEAK}$  &          &  & $Z_{WEIGHT}$          &\\
\hline
\Pzpdf\ Code       & $\frac{\sigma_{IQR}}{(1+z)}$ & median  & \multicolumn{1}{|p{0.75cm}|}{\centering outlier \\fraction} & $\frac{\sigma_{IQR}}{(1+z)}$ & median & \multicolumn{1}{|p{0.75cm}|}{\centering outlier \\fraction}\\
\hline
\annz     & 0.0270  &  0.00063  & 0.044      & 0.0244  &  0.000307  & 0.047  \\
\bpz       & 0.0215  & -0.00175  & 0.035      & 0.0215  & -0.002005  & 0.032 \\
\delight   & 0.0212  & -0.00185  & 0.038      & 0.0216  & -0.002158  & 0.038 \\
\eazy      & 0.0225  & -0.00218  & 0.034      & 0.0226  & -0.003765  & 0.029 \\
\flexzboost& 0.0154  & -0.00027  & 0.020      & 0.0148  & -0.000211  & 0.017 \\
\gpz       & 0.0197  & -0.00000  & 0.052      & 0.0195  &  0.000113  & 0.051 \\
\lephare   & 0.0236  & -0.00161  & 0.058      & 0.0239  & -0.002007  & 0.056 \\
\metaphor  & 0.0264  &  0.00000  & 0.037      & 0.0262  &  0.001333  & 0.048 \\
\cmnn        & 0.0184  & -0.00132  & 0.035      & 0.0170  & -0.001049  & 0.034 \\
\skynet    & 0.0219  & -0.00167  & 0.036      & 0.0218  &  0.000174  & 0.037 \\
\tpz       & 0.0161  &  0.00309  & 0.033      & 0.0166  &  0.003048  & 0.031 \\
\hline
\trainz   & 0.1808  &  -0.2086  & 0.000  & 0.2335  & 0.022135  & 0.000\\
\end{tabular}
\end{center}
\end{table*}

The intrinsic scatter, bias, and catastrophic outlier rate are given in Table~\ref{tab:pointestimates}.
Perhaps unsurprisingly, performance under these metrics largely tracks that of the metrics of Section~\ref{sec:metrics} of the \pzpdf s from which the point estimates were derived.
All twelve codes perform at or near the goals of the \lsst\ Science Requirements Document\footnote{available at: \url{http://ls.st/srd}} and \citet{Graham:17}, which is encouraging if not unexpected for $i < 25.3$.

\section*{Data Availability}
The data underlying this article were provided by the makers of the Buzzard Flock simulation group by permission.  Data will be shared on request to the corresponding authors upon permission of the Buzzard Flock simulation group.  For updated and larger datasets, requests can be made directly to the authors of the Buzzard Flock simulations \citep{DeRose:19}.  All analysis and plotting scripts used in the writing of this paper are available on GitHub in the \texttt{metric\_scripts} and \texttt{plotting\_scripts} directories of the following repository: \url{https://github.com/LSSTDESC/PZDC1paper}.

\bibliography{main,lsstdesc}

\section*{Affiliations}
$^{1}$ Department of Physics, University of California, One Shields Ave., Davis, CA, 95616, USA\\
$^{2}$ German Centre of Cosmological Lensing, Ruhr-Universitaet Bochum, Universitaetsstra{\ss}e 150, 44801 Bochum, Germany\\
$^{3}$ Center for Cosmology and Particle Physics, New York University, 726 Broadway, New York, 10003, USA\\
$^{4}$ Department of Physics, New York University, 726 Broadway, New York, 10003, USA\\
$^{5}$ School of Physics, University of Science Malaysia, 11800 USM, Pulau Pinang, Malaysia\\
$^{6}$ Department of Physics and Astronomy, University College London, Gower Street, London WC1E 6BT, UK\\
$^{7}$ King Abdulaziz City for Science and Technology, Riyadh 11442, Saudi Arabia\\
$^{8}$ Information Engineering, Parks Road, Oxford, OX1 3PJ, UK\\
$^{9}$ INAF-Astronomical Observatory of Capodimonte, Salita Moiariello 16, 80131 Napoli, Italy\\
$^{10}$ Department of Physics E. Pancini, University Federico II, via Cinthia 6, I-80126, Napoli, Italy\\
$^{11}$ Laboratoire Univers et Particules de Montpellier, Universit\'e de Montpellier, CNRS, Montpellier, France\\
$^{12}$ DIRAC Institute and Department of Astronomy, University of Washington, Box 351580, U.W., Seattle WA 98195, USA\\
$^{13}$ Santa Cruz Institute for Particle Physics, Santa Cruz, CA 95064, USA\\
$^{14}$ Berkeley Center for Cosmological Physics, Department of Physics, University of California, Berkeley CA 94720\\
$^{15}$ Lawrence Berkeley National Laboratory, 1 Cyclotron Road, Berkeley, CA 94720, USA\\
$^{16}$ Kavli Institute for Particle Astrophysics and Cosmology and Department of Physics, Stanford University, Stanford, CA 94305, USA\\
$^{17}$ Department of Particle Physics and Astrophysics, SLAC National Accelerator Laboratory, Stanford, CA 94305, USA\\
$^{18}$ Department of Statistics \& Data Science, Carnegie Mellon University, 5000 Forbes Avenue, Pittsburgh, PA 15213, USA\\
$^{19}$ Dunlap Institute for Astronomy \& Astrophysics, University of Toronto, 50 St. George Street, Toronto, ON M5S 3H4, Canada\\
$^{20}$ Department of Physics and Astronomy, Rutgers, The State University of New Jersey, 136 Frelinghuysen Road, Piscataway, NJ 08854-8019 USA\\
$^{21}$ Astrophysics, Department of Physics, University of Oxford, Denys Wilkinson Building, Keble Road, Oxford, OX1 3RH, UK\\
$^{22}$ Department of Physics and Astronomy, University of the Western Cape, Bellville 7535, South Africa\\
$^{23}$ Argonne National Laboratory, Lemont, IL 60439, USA\\
$^{24}$ Department of Physics and Astronomy and the Pittsburgh Particle Physics, Astrophysics and Cosmology Center (PITT PACC), University of Pittsburgh, Pittsburgh, PA 15260, USA\\
$^{25}$ Department of Physics, Stanford University, 382 Via Pueblo Mall, Stanford, CA 94305, USA\\
$^{26}$ SLAC National Accelerator Laboratory, Menlo Park, CA, 94025, USA\\
$^{27}$ Department of Statistics, Federal University of Sao Carlos, Sao Carlos, Brazil\\
$^{28}$ External collaborator\\

\end{document}